\documentclass[iop, twocolappendix]{emulateapj}
\usepackage{graphicx}
\usepackage{txfonts}
\usepackage{xcolor}
\usepackage[colorlinks, citecolor = blue]{hyperref}
\usepackage{epsfig}
\usepackage{natbib}

\shorttitle{ALMA reveals large overdensity and strong clustering of galaxies in quasar environments at $z\sim4$}
\shortauthors{Garc\'ia-Vergara et al.}

\begin{document}

\title{ALMA reveals large overdensity and strong clustering of galaxies in quasar environments at $z\sim4$}

\author{Cristina Garc\'ia-Vergara\altaffilmark{1}}
\author{Matus Rybak\altaffilmark{1,2}}
\author{Jacqueline Hodge\altaffilmark{1}}
\author{Joseph F. Hennawi\altaffilmark{3}}
\author{Roberto Decarli\altaffilmark{4}}
\author{Jorge Gonz\'alez-L\'opez\altaffilmark{5,6}}
\author{Fabrizio Arrigoni-Battaia\altaffilmark{7}}
\author{Manuel Aravena\altaffilmark{5}}
\author{Emanuele P. Farina\altaffilmark{7}}

\altaffiltext{1}{Leiden Observatory, Leiden University, Niels Bohrweg 2, 2333 CA Leiden, the Netherlands}
\altaffiltext{2}{Faculty of Electrical Engineering, Mathematics and Computer Science, TU Delft, the Netherlands}
\altaffiltext{3}{Department of Physics, University of California, Santa Barbara, CA 93106, USA}
\altaffiltext{4}{INAF – Osservatorio di Astrofisica e Scienza dello Spazio di Bologna, Via Gobetti 93/3, 40129 Bologna, Italy}
\altaffiltext{5}{N\'ucleo de Astronom\'ia de la Facultad de Ingenier\'ia y Ciencias, Universidad Diego Portales, Av. Ej\'ercito Libertador 441, Santiago, Chile}
\altaffiltext{6}{Las Campanas Observatory, Carnegie Institution of Washington, Casilla 601, La Serena, Chile}
\altaffiltext{7}{Max-Planck-Institut f\"ur Astrophysik, Karl-Schwarzschild-Str 1, 85748 Garching bei M\"unchen, Germany}

\begin{abstract}
We present an Atacama Large Millimeter/submillimeter Array (ALMA) survey of CO(4--3) line emitting galaxies in 17 quasar fields at $z\sim4$ aimed to perform the first systematic search of dusty galaxies in high$-z$ quasar environments. Our blind search of galaxies around the quasars results in 5 CO emitters with S/N$\geq5.6$ within a projected radius of $R\lesssim1.5\,h^{-1}\,$cMpc and a velocity range of $\rm\Delta\,v=\pm1000\,$km\,s$^{-1}$ around the quasar. In blank fields, we expect to detect only 0.28 CO emitters within the same volume, implying a total overdensity of $17.6^{+11.9}_{-7.6}$ in our fields, and indicating that quasars trace massive structures in the early universe. We quantify this overdensity by measuring the small-scale clustering of CO emitters around quasars, resulting in a cross-correlation length of $r_{\rm 0,QG}=8.37^{+2.42}_{-2.04}\,h^{-1}\,$cMpc, assuming a fixed slope $\gamma=1.8$. This contradicts the reported mild overdensities (x1.4) of Lyman alpha emitters (LAEs) in the same fields at scales of $R\lesssim7\,h^{-1}\,$cMpc which is well described by a cross-correlation length 3 times lower than that measured for CO emitters. We discuss some possibilities to explain this discrepancy, including low star formation efficiency, and excess of dust in galaxies around quasars. Finally, we constrain, for the first time, the clustering of CO emitters at $z\sim4$, finding an auto-correlation length of $r_{\rm 0,CO}=3.14\pm1.71\,h^{-1}\,$cMpc (with $\gamma=1.8$). Our work together with the previous study of LAEs around quasars traces simultaneously the clustering properties of both optical and dusty galaxy populations in quasars fields, stressing the importance of multi-wavelength studies, and highlighting important questions about galaxy properties in high$-z$ dense environments. 
\end{abstract}


\section{Introduction}

Theoretical studies and simulations of the formation and evolution of large-scale structure in the universe have been largely developed in the last decades, but observational constraints in this field remain one of the major challenges for astrophysicists. In the standard current paradigm, the cosmological model is parametrized by the Lambda cold dark matter ($\Lambda$CDM) model, and the structures grow hierarchically through gravitational instability \citep[e.g.][]{Dodelson2003,Padmanabhan2006,Schneider2015}, forming a filamentary structure of dark matter and galaxies. In this framework, the most massive galaxies in the early universe are placed in the most massive dark matter halos \citep{Springel2005,Angulo2012}, tracing the most massive structures. Since the dark matter halo mass can be directly related to the clustering of a population \citep{Cole1989,Mo1996,Sheth1999}, the generic prediction is that massive galaxies should have a large auto-correlation function. 

The population with the largest auto-correlation function in the early universe are quasars, and thus they represent key objects to test structure formation models at early epochs. Quasars are accreting supermassive black holes (SMBHs), and are some of the most luminous sources in the universe. They are observed to peak at $z\sim2-3$ \citep[e.g.][]{Richards2006} and have been detected up to $z=7.5$ \citep{Banados2018,Yang2020}. The quasar auto-correlation has been measured as a function of redshift up to $z\sim4$, and it has been found to rise steadily over the redshift range $0\lesssim z\lesssim2$ \citep{Croom2005,Myers2006,Porciani2006,Shen2007,daAngela2008} with a clustering similar to that of local galaxies, and to strongly increase between $z\sim2$ and $z\sim4$ \citep{Shen2007,White2012,Eftekharzadeh2015,Timlin2018}. 

Additionally, the quasar auto-correlation function has been shown to get progressively steeper on small ($R\lesssim40\,h^{-1}\rm pkpc$) scales at $z<3$ \citep{Hennawi2006}, and the small-scale clustering has been shown to also increase from $z\sim3$ to $z\sim4$ \citep{Shen2010}. The measurements of $z\sim4$ quasar clustering result in an auto-correlation length $r_0=24.3\pm2.4\,h^{-1}\rm cMpc$ (at a fixed slope $\gamma=2.0)$ implying that quasars are the most clustered population at this redshift and that they inhabit very massive ($M_{\rm halo} > 6 \times 10^{12} M_{\odot}\,h^{-1}$) dark matter halos \citep{Shen2007}, in agreement with theoretical predictions \citep{Springel2005,Angulo2012}. These results imply that $z\sim4$ quasars are tracers of massive structures, and thus they should be surrounded by a large overdensity of galaxies. 

Several efforts have been made to detect such overdensities in quasars environments, but the results reveal a contradictory and confusing picture. Most of the quasar environment studies at $z\gtrsim4$ have been performed at optical wavelengths, and they have been aimed at detecting overdensities of Lyman break galaxies (LBGs) or Lyman alpha emitters (LAEs) around individual quasars. Some of the studied fields show overdensities of galaxies \citep{Stiavelli2005,Zheng2006,Kashikawa2007,Kim2009,Utsumi2010,Capak2011,Swinbank2012,Morselli2014,Adams2015,Farina2017,Balmaverde2017,Kikuta2017,Ota2018}
 whereas others exhibit a similar number density of galaxies compared with blank fields \citep{Kim2009,Willott2005,Banados2013,Husband2013,Simpson2014,Mazzucchelli2017,Goto2017}. It thus remains unclear whether $z\gtrsim4$ quasars reside in overdensities as would be implied by their strong auto-correlation. The contradictory results obtained so far might be a consequence of heterogeneous methodologies, depth, and probed physical scales, together with the low number statistics, and the large cosmic variance expected to affect these studies \citep{Garcia-Vergara2019,Buchner2019}.

One strategy that has been recently used to overcome the low number statistics and high cosmic variance is to characterize quasar environments by targeting a larger number of fields and measuring the quasar-galaxy cross-correlation function. Using this technique, \citet{Garcia-Vergara2017} demonstrate that LBGs at scales $R\lesssim9\,h^{-1}\rm cMpc$ are strongly clustered around $z\sim4$ quasars with a cross-correlation length $r_0=8.83^{+1.39}_{-1.51}\,h^{-1}\rm cMpc$ (at a fixed slope $\gamma=2.0)$ and an overall overdensity of 1.5. They find a very good agreement with the expectations assuming a deterministic bias model, in which quasars and galaxies trace the same underlying dark matter distribution. \citet{Garcia-Vergara2019} studied the environments of 17 bright quasars at $z\sim4$ at scales of $R\lesssim7\,h^{-1}\rm cMpc$ traced by LAEs (with $\rm log(L_{\rm Ly\alpha}[erg/s])\geq 42.61$) selected using narrow band imaging. They find an average LAE overdensity around quasars of 1.4 for the full sample, and a cross-correlation length of $r_0=2.78^{+1.16}_{-1.05}\,h^{-1}\rm cMpc$ (at a fixed slope $\gamma=1.8)$. Although the cross-correlation was found to be consistent with a power-law shape, indicative of a concentration of LAEs centered on quasars, they find 2.1 times fewer LAEs than the expectation computed by assuming a deterministic bias model. 

This work contradicts the findings of recent deeper ($\rm log(L_{\rm Ly\alpha}[erg/s])\geq 42.1$) observations of 27 $z=3-4.5$ quasars environments performed with Multi Unit Spectroscopic Explorer (MUSE; \citealt{Bacon2010}) on the Very Large Telescope (VLT) that revealed a large presence of LAEs \citep{Fossati2021}. They find an overdensity of 4.1 at scales $R\lesssim0.6\,h^{-1}\rm cMpc$, and they measure a cross-correlation length $r_0=3.15^{+0.36}_{-0.40}\,h^{-1}\rm cMpc$ at a fixed slope $\gamma=1.8$\footnote{The $r_0$ value published in \citet{Fossati2021} has been recently re-computed. Here we quote the most updated $r_0$ value, which was kindly provided by the authors by private communication.}. 

The aforementioned works statistically demonstrate the existence of galaxy overdensities concentrated around quasars at $z\sim4$, but they are still contradictory in the detection of the ubiquitous large galaxy concentration expected based on the theoretical deterministic bias model. Various possibilities can be cited to explain this discrepancy. First, most of the previous studies rely on quasar redshifts determined from rest-frame UV emission lines which are known to be offset from the quasar systemic redshifts \citep[e.g.][]{Richards2002,Shen2007,Shen2016,Coatman2017}. If these offsets are large, it would imply that the search for galaxies has actually been performed at higher or lower radial comoving distances from the quasar, where the clustering signal is vanishingly small, resulting in a lower number density of galaxies, and thus the expected large overdensities would not be detectable. 

A second possibility is that the complex physical processes associated with quasar radiation affect the visibility of LAEs in their vicinity. On one hand ionizing radiation coming from the quasar could suppress star formation in nearby galaxies \citep[e.g.][]{Francis2004,Bruns2012} making the Ly$\alpha$ line emission and UV continuum less intense, and implying a lower number of detectable galaxies in quasar environments. On the other hand, the ionizing radiation from the quasar could induce Ly$\alpha$ fluorescent emission \citep{Cantalupo2012}, resulting in an increase of the LAE number density. These two effects work in opposite directions and disentangling them becomes extremely difficult unless the environments are traced by a galaxy population less affected by such feedback effects. Note however that only small-scale environment studies would be more impacted by the mentioned effects since the ionizing radiation is not expected to extend up to large scales, at which most previous quasar environment studies have been performed.   

Finally, galaxies in quasar environments could be more dusty and thus the optical/UV emission is slightly obscured, making them undetectable at optical wavelengths. This would explain why only deep LAE studies reveal large overdensities \citep[e.g.][]{Fossati2021}, while shallower observations miss a large fraction of them \citep[e.g.][]{Garcia-Vergara2019}. Submillimeter observations, which are less affected by dust, would be required to explore this possibility. Indeed, some serendipitous Atacama Large Millimeter/submillimeter Array (ALMA) detections of dusty galaxies have been recently reported in high$-z$ quasars environments, on scales of $R<0.5\rm\, cMpc$  \citep{Decarli2017,Trakhtenbrot2017,Venemans2020}, suggesting a possible strong clustering of dusty galaxies around quasars. However, the quasar-galaxy cross-correlation function has not been constrained.

Summarizing, incomplete sampling of the galaxy population around quasars could be responsible for the contradictory results obtained so far. Systematic and multi-wavelength studies are still required to statistically quantify galaxy overdensities of different galaxy populations in $z\sim4$ quasar environments. 
 
In this paper, we present an ALMA survey of CO(4--3) line emitters in the environs of 17 quasars at $z\sim4$, the same quasar sample recently studied for LAE overdensities by \citet{Garcia-Vergara2019}, to constrain the small-scale ($R\lesssim1.5\,h^{-1}\rm cMpc$) clustering of CO emitters, and ultimately trace simultaneously the clustering properties of both optical and dusty galaxy populations around quasars. This is the first quasar sample observed with ALMA for environmental studies purposes and thus provides an opportunity to statistically trace  -- for the first time -- the overdensity and clustering of dusty galaxies around high$-z$ quasars.

Additionally, our observations provide a more precise measurement of the quasar redshift, previously determined from rest-frame UV emission lines, which is important for clustering analyses. We note that with our observations we could also trace the number density of continuum sources; however, this information is not useful for clustering analysis. Given the roughly constant flux density of galaxies at submillimeter wavelengths across the redshift range $1\lesssim z\lesssim7$ \citep{Blain2002}, we would not distinguish between foreground and background sources, and therefore all the detected sources would be included in the clustering computation. The clustering signal would thus be integrated over large comoving volumes, which would strongly dilute the real clustering signal in the close environment of quasars. 

The focus of the present paper is on the CO(4--3) emitter number counts in quasar environments, while two forthcoming papers (Garc\'ia-Vergara et al. in preparation) analyze the quasar host galaxy properties and the multi-wavelength properties of galaxies in quasar fields, respectively.

The outline of the paper is as follows. In \S~\ref{sec:data} we describe the targeted quasar sample, the available ancillary data in the fields, the ALMA observations, and data reduction. In \S~\ref{sec:catalog} we describe the detection and selection of CO(4--3) emitting galaxies and present the final catalog. The galaxy number counts and clustering analysis are presented in \S~\ref{sec:clustering}. We discuss and interpret our results in \S~\ref{sec:discussion}, and we summarize in \S~\ref{sec:summary}. The Appendix discusses the impact of several factors on our clustering measurement. Throughout the paper we adopt a $\Lambda$CDM cosmological model $h=0.7$, $\Omega_{m}=0.30$, $\Omega_{\Lambda}=0.70$, and $\sigma_8=0.8$ which is consistent with \citet{Planck2018}. Comoving and proper Mpc are denoted as ``cMpc" and ``pMpc", respectively. Magnitudes are given in the AB system \citep{Oke1974,Fukugita1995}.

\section{Observations and Data Reduction}
\label{sec:data}

\subsection{Quasar sample}
\label{ssec:quasars}
The quasar sample used in this study is the same sample presented in \citet{Garcia-Vergara2019} who observed 17 quasar fields at optical wavelengths to search for LAEs in $z\sim4$ quasar environments. We refer the reader to the mentioned work for details of the quasar selection strategy. Briefly, the 17 quasars were selected from the Sloan Digital Sky Survey (SDSS; \citealt{York2000}) and the Baryon Oscillation Spectroscopic Survey (BOSS; \citealt{Eisenstein2011,Dawson2013}) quasar catalog \citep{Paris2014} to lie within a redshift window of $z\sim3.862-3.879$ (corresponding to $\rm \Delta v = 1066 \,km\,s^{-1}$ at $z=3.87$). This redshift window was chosen based on the coverage of the optical narrow-band (NB) filter used to detect LAEs. The quasar redshifts were determined based on one or more rest-frame UV emission lines (SIV $\lambda$ 1396, CIV $\lambda$ 1549, and CIII] $\lambda$ 1908) and using the calibration of emission-line shifts from \citet{Shen2007} to estimate the systemic redshift. All the quasars have $1\sigma$ redshift uncertainties $\lesssim\rm 800\,km\,s^{-1}$. We tabulate the quasar positions, redshifts, and $i$-band magnitudes in Table~\ref{table:quasar}.

\begin{table}
\caption{Optical properties of the targeted quasars. \label{table:quasar}}
\centering
\begin{tabular}{lccc c}
\hline 
Target ID & RA & Dec & Redshift & $i$ mag\\
(SDSS) & (J2000) & (J2000) &   &  \\
\hline
J0040+1706 & 00:40:17.426 &  +17:06:19.78 &   3.873  $\pm$ 0.008&18.91\\
J0042-1020 &  00:42:19.748 &  -10:20:09.53 &  3.865    $\pm$ 0.012&18.57\\
J0047+0423 &  00:47:30.356 &  +04:23:04.73 &   3.864  $\pm$ 0.008&19.94\\
J0119-0342 & 01:19:59.553 &  -03:42:16.51 &  3.873    $\pm$ 0.013&20.49\\
J0149-0552 & 01:49:06.960 &  -05:52:18.85 &  3.866    $\pm$ 0.013&19.80\\
J0202-0650 & 02:02:53.765 &  -06:50:44.54 &  3.876    $\pm$ 0.008&20.64\\
J0240+0357 & 02:40:33.804 &  +03:57:01.59 &    3.872 $\pm$ 0.012&20.03\\
J0850+0629 & 08:50:13.457 &	+06:29:46.91   & 3.875 $\pm$ 0.013&20.40\\
J1026+0329 & 10:26:32.976  & +03:29:50.63  &   3.878  $\pm$ 0.008&19.74\\
J1044+0950 & 10:44:27.798  & +09:50:47.98  &   3.862  $\pm$ 0.012&20.52\\
J1138+1303 & 11:38:05.242  & +13:03:32.61  &   3.868  $\pm$ 0.008&19.10\\
J1205+0143 & 12:05:39.550  & +01:43:56.52  &   3.867  $\pm$ 0.008&19.37\\
J1211+1224 & 12:11:46.935   & +12:24:19.08    &  3.862   $\pm$ 0.008&19.97\\ 
J1224+0746 &  12:24:20.658   &+07:46:56.33   &  3.867   $\pm$ 0.008&19.08\\
J1258-0130 & 12:58:42.118   &-01:30:22.75   & 3.862     $\pm$ 0.008&19.58\\ 
J2250-0846 & 22:50:52.659  & -08:46:00.22  & 3.869    $\pm$ 0.012&19.44\\
J2350+0025 &  23:50:32.306  & +00:25:17.23  &   3.876  $\pm$ 0.012&20.61\\
\hline
\end{tabular}
\tablecomments{Quasar positions are determined from optical images (SDSS/BOSS quasar catalog; \citealt{Paris2014}), redshifts are based on one or more rest-frame UV emission lines and calibrated to estimate the systemic redshift (see \S~\ref{ssec:quasars}), and magnitudes correspond to the $i$-band magnitudes from SDSS.}
\end{table}

\subsection{Ancillary data}
\label{ssec:obs_opt}
The 17 quasar fields were previously observed (Program ID: 094.A-0900) using the FOcal Reducer and low dispersion Spectrograph 2 (FORS2, \citealt{Appenzeller1992}) instrument on the VLT in 30 hours of observing time. Optical deep observations were acquired using the narrow band HeI ($\lambda_{\rm c} = 5930$\AA, $\rm FWHM = 63$\AA) and the broad bands $g_{\rm HIGH}~(\lambda_{\rm c} = 4670$\AA, hereafter $g$) and $R_{\rm SPECIAL}~(\lambda_{\rm c} = 6550$\AA, hereafter $R$) to detect LAEs around the quasars. The total area observed per field was $6.8\times6.8$ arcmin$^{2}$, with an image pixel scale of $\rm 0.25 \arcsec/pix$. The median of the $5\sigma$
limiting magnitude reached in the fields was 24.45, 25.14 and 25.81 for the HeI, $R$ and $g$ images respectively. 

A catalog of 25 LAEs within $R\lesssim7\,h^{-1}\rm cMpc$ in the quasar fields is available \citep{Garcia-Vergara2019}. LAEs were selected based on a color selection criteria such that they have a (rest-frame) EW$_{\rm Ly\alpha}>28$\AA\, and a solid detection of ${\rm S\slash N}\ge 5.0$ in the HeI band. All the LAEs lie approximately within $\rm \pm1598 \,km\,s^{-1}$ (corresponding to $\pm18.7\,\rm cMpc$ at $z=3.78$) from the quasar systemic redshift, as given by the coverage of the used NB filter. Only two out of 25 LAEs lie within a radius of $59\arcsec$ from the central quasar, which is the area covered by the ALMA observations presented in this study.

\subsection{ALMA Band~3 observations}
\label{ssec:obs}
Our observations were carried out during ALMA Cycle~7, as part of the observing program \#2019.1.00411.S (PI: C. Garcia Vergara). The data were taken in Band~3 (84-116 GHz) between 2019 October 4 and 2020 March 19. Most targets were observed in a single scheduling block with 45-50 minutes on-source, except for SDSS J0040+1706, SDSSJ1138+1303, SDSSJ1205+0143, and SDSSJ1224+0746 which were observed in two scheduling blocks. 

The 17 pointings were centered at the optical coordinates of each quasar (specified in Table~\ref{table:quasar}). The spectral setup consists of four spectral windows (SPWs) with a 1.9~GHz bandwidth per SPW. One of the SPWs (hereafter SPW0) was centered on the expected observed frequency of the CO(4--3) emission line from the central quasar, given the quasar systemic redshifts reported in Table~\ref{table:quasar}. The 1.9~GHz bandwidth correspond to $\rm \Delta v\sim 6000\,km\,s^{-1}$ (or $\Delta z=0.098$) at the observed frequencies in SPW0. The SPW0 was placed into the lower sideband and configured with a spectral channel width of 7.8125~MHz (equivalent to $\rm \sim25\,km\,s^{-1}$ at the observed frequency).

The remaining three SPWs were located at higher frequencies, with one of them (SPW1) adjacent to SPW0 and the other two (SPW2 and SPW3) in the upper sideband (at $\rm \sim12 GHz$ from the lower sideband). SPWs 1, 2, and 3 were configured with a 15.625~MHz channel width. 

Most observations were taken in the C43-4 array configuration with a maximum baseline length of 783\,m, but several observations were taken in a more extended configuration, with a maximum baseline length of 1231 or 2517\,m. Table~\ref{tab:obs} summarizes the details of individual observing blocks. The shortest baseline was always 15\,m, and the maximum recoverable scale ranged between 10 and 14\arcsec. The number of 12-m antennas used ranged between 32 and 49. 

The primary beam FWHM is 66\arcsec at 94.5~GHz, and the typical size of the synthesized beam is $1.4\arcsec \times 1.2\arcsec$ (see Table~\ref{tab:obs}).

\subsection{Imaging}
\label{ssec:imaging}
All data processing and imaging was performed using \textsc{Casa} (Common Astronomy Software Applications package, \citealt{McMullin2007}), version 5.6.1. We process the data using the standard ALMA pipeline, however, some datasets required further manual flagging of problematic antennas. Specifically, we flagged the antennas DA60 and DA58 in J1258-0130, DV04 and DV11 in J0240+0357, DV19 in J1138+1303 (2020 March 19 observation), and DV17 for J1044+0950. The sources were too faint for a successful self-calibration.

For each quasar field, we used \textsc{Casa}'s task \texttt{tclean} to create both dirty-image and clean spectral cubes at native frequency resolution. We use the H\"ogbom cleaning algorithm and re-calculate the noise per visibility (setting \texttt{fastnoise = False}). All images were created using natural weighting to maximize the surface brightness sensitivity. For the cleaned images, we used a stopping threshold of $1.5\sigma$ without manual masking. We set the pixel size to 0.2\arcsec, which ensures that the synthesized beam is sufficiently sub-sampled. Note that the line search was performed on the dirty cubes, and the clean cubes are only used to extract the detected sources (see \S~\ref{ssec:LSA}).

J0240+0357 was observed with a rather sparse uv-plane coverage, resulting in a noticeably non-Gaussian dirty beam with large sidelobes. To improve the image quality, we re-imaged this source using an outer Gaussian taper of 1.5\arcsec.

We use the dirty spectral cubes created from SPW0 (the cube at the expected observed frequency of the quasar emission line) without a primary beam correction to compute the rms noise per 7.813 MHz channel throughout the scanned frequency range in a rectangular region that covers most of the region within the primary beam FWHM. The rms noise varies only slightly per channel, (typically $\sim3\%$), so we compute the median rms noise measured in all the channels of the spectral cube for each field and report these values in Table~\ref{tab:obs}. 

Finally, we create a continuum image by combining all four SPWs. For each quasar field, we compute the rms noise in the dirty continuum images (without primary beam correction) over the same region as for the SPW0 cube and tabulate the values in Table~\ref{tab:obs}. The achieved rms is in agreement with expectations from the ALMA sensitivity calculator. Note that in our analysis, we do not subtract the continuum from the data.

\begin{table*}
\caption{ALMA observations and imaging summary.} 
\centering
\begin{tabular}{llcccccc}
\hline
Target ID & Dates observed & $N_\mathrm{ant}$ & Max baseline & Beam size (PA)& $\sigma_\mathrm{cont}$& $\sigma_\mathrm{SPW0}$ &log$L'_{\rm CO(4-3)}$ \\
& & & [m] & [arcsec$\times$arcsec], [deg] & $\mu$Jy/beam & mJy/beam&[$\rm K\,km\,s^{-1}pc^2$]\\
\hline
J0040+1706 & 2020 Mar 17 & 32 & 1231~m & 1.21$\times$1.00 (36) & 12.8&0.32&9.78\\
& 2020 Mar 18 & 39 & 1231~m & & \\
J0042-1020 & 2019 Oct 19 & 49 & 783~m & 1.42$\times$1.19 (-85) & 8.9&0.23&9.64\\
J0047+0423 & 2019 Oct 14 & 48 & 783~m & 1.38$\times$1.26 (-37) & 9.0&0.24&9.65\\
J0119-0342 & 2019 Oct 19 & 43 & 783~m & 1.41$\times$1.22 (-78) & 8.4&0.23&9.64\\
J0149-0552 & 2019 Oct 19 & 49 & 783~m & 1.39$\times$1.19 (-65) & 8.8&0.24&9.66\\
J0202-0650 & 2019 Oct 19 & 49 & 783~m & 1.49$\times$1.17 (-62) & 8.5&0.23&9.64\\
& 2019 Oct 14$^\dagger$ & 43 & 783~m & & &\\
J0240+0357 & 2019 Oct 4 & 48 & 2517~m & 1.24$\times$0.91 (-53) & 14.1&0.35&9.82\\
J0850+0629 & 2020 Mar 2 & 44 & 783~m & 1.38$\times$1.22 (49) & 10.1&0.25&9.67\\
J1026+0329 & 2019 Oct 19 & 46 & 783~m & 1.37$\times$1.24 (-58) & 9.8&0.23&9.63\\
J1044+0950 & 2020 Mar 2 & 43 & 783~m & 1.41$\times$1.30 (23) & 9.5&0.23&9.64\\
J1138+1303 & 2020 Mar 16 & 43 & 783~m & 1.30$\times$1.17 (42) & 9.3&0.24&9.66\\
& 2020 Mar 19 & 45 & 1231~m & &&\\
J1205+0143 & 2019 Oct 19 & 46 & 783~m & 1.33$\times$1.23 (-76) & 8.5&0.23&9.64\\
& 2020 Mar 3 & 43 & 784~m & &&\\
J1211+1224 & 2020 Mar 15 & 43 & 783~m & 1.47$\times$1.20 (-32) & 11.8&0.29&9.74\\
J1224+0746 & 2020 Mar 14 & 40 & 783~m & 1.38$\times$1.24 (-40) & 8.7&0.23&9.64\\
 & 2020 Mar 15 & 40 & 783~m & &&\\
J1258-0130 & 2020 Mar 3 & 44 & 783~m & 1.46$\times$1.18 (-46) & 10.7&0.30&9.75\\
J2250-0846 & 2019 Oct 18 & 47 & 783~m & 1.39$\times$1.18 (-86) & 10.4&0.25&9.67\\
J2350+0025 & 2019 Oct 14 & 48 & 783~m & 1.35$\times$1.24 (-50) & 9.0&0.25&9.67\\ 
\hline
\multicolumn{4}{l}{$\dagger$ semi-pass observation}
\end{tabular}
\tablecomments{For each target, we list the dates of observations, the number of antennas used, and the baseline range. The synthesized beam sizes are for naturally weighted images for the final data products. We include the rms noise measured in the dirty continuum images and the rms noise per channel (channel width of 7.8125~MHz, equivalent to $\rm \sim25\,km\,s^{-1}$ at the observed frequency) measured in the dirty spectral cube created from SPW0 (whihc is targeting the expected CO(4--3) emission). We also report the limiting luminosity of the CO line at $5.6\sigma$ (the S/N threshold used in this work), assuming a line width of $331\rm\,km\,s^{-1}$ (see details in \S~\ref{ssec:cross}). Since the rms is measured in the non-primary beam corrected image, the reported limiting luminosity corresponds to the center of the pointing for each field, but this increases with increasing radius.\label{tab:obs}}
\end{table*}

\section{Emission line catalog}
\label{sec:catalog}
In this study, we focus on the measurement of the small-scale clustering of CO(4--3) emission lines around the central quasar. Therefore, we limit our emission line search to the spectral cube created from SPW0, where the central quasar is expected to be located. The procedure and catalog presented in this section only contain sources detected in this spectral cube, which cover $\rm \pm3000\,km\,s^{-1}$ around the quasar. At the end of this section, we briefly present the detection of the quasars, and the comparison between their optical and ALMA redshifts (see \S~\ref{ssec:quasars}). We present further details about the quasars in a forthcoming paper (Garc\'ia-Vergara et al. in preparation). 

 \subsection{Line search algorithm}
\label{ssec:LSA}
We perform a blind search for emission lines in the quasar fields (i.e.\,on the 17 spectral cubes extracted from the SPW0). Since we do not expect to detect extremely bright sources in our cubes, we prefer to use the ``dirty" cubes to perform the line search, allowing us to preserve the intrinsic properties of the noise in the cubes. We use \textsc{findclumps}, an IRAF-based routine originally developed for the blind search of CO lines performed in the ALMA spectroscopic survey in the Hubble Ultra Deep Field (ASPECS; \citealt{Walter2016,Decarli2019}), a 3D survey of gas and dust in distant galaxies. We refer the reader to these works for further details about the algorithm implemented in the \textsc{findclumps} routine. Briefly, \textsc{findclumps} performs a top-hat convolution using $N_{\rm chn}$ consecutive frequency channels of the cube to create a combined image, with $N_{\rm chn}$ ranging from 3 up to 19 (corresponding to $\rm 74-470\,km\,s^{-1}$ at $z=3.87$) in steps of 2 channels. For each combined image, the rms is computed\footnote{We use a rectangular region that covers most of the region within the primary beam FWHM, since the primary beam pattern should be very well defined, and thus the image before the primary beam correction should be flat, without spurious features, and with (mostly) Gaussian noise.}, and a search for sources is performed using SExtractor \citep{Bertin1996} resulting in one source catalog. The signal-to-noise (S/N) of each source in this catalog is computed using the peak flux of the source and the rms computed for the combined image. We only keep sources with S/N$\geq3.5$ in the catalog. The search is performed not only within the primary beam FWHM but over the entire area covered by our observations which is given by a circular area with a radius $59\arcsec$ (at which the telescope sensitivity is 10\% of the maximum).

Since the source search is performed repetitively using different numbers of channels, we obtain several individual source catalogs, and we combine these in one final catalog per cube; however, the resulting final catalog contains duplications. For example, if the FWHM of an emission line is equivalent to 19 channels, the algorithm will detect such a line with the maximum S/N in the combined image created by using 19 consecutive frequency channels. However, that source may also be detected in the combined image created by using fewer consecutive channels, although with lower S/N since in this case only part of the total flux is encompassed in the combined image. 

Since the S/N of a line is maximized when detected combining a number of channels equivalent to the exact width of the line, we look for all the sources that fall within 1.5\arcsec (which is similar to the size of the synthesized beam in our images) and $\rm 650\,km\,s^{-1}$ (0.21 GHz at $z=3.87$), and we only keep in our final catalogs the source with the highest S/N. This procedure effectively removes the duplications, but we could also be missing some real close pair of sources (for example mergers). We repeat this line search process on all our cubes, resulting in 17 source catalogs. 

\subsection{Fidelity}
\label{ssec:fidelity}
The assessment of the reliability of the sources in our catalogs is crucial to distinguish real detections from spurious detections only caused by noise peaks exceeding our S/N threshold. The reliability of a line can be determined by exploring the noise properties of each cube. Specifically, we apply the same line search algorithm described in \S~\ref{ssec:LSA} to the negative cubes, and we create final catalogs containing (unphysical) negative lines detected with S/N$\geq3.5$. Since negative lines are only produced by noise fluctuations, this catalog can be used to determine the probability that a positive line detection is due to noise as a function of their S/N. Note however that this probability also depends on the line width, such that at a fixed S/N, a source with a wider line has a higher probability to be real than a source with a narrower line (see a detailed discussion in \citealt{Gonzalez-lopez2019}). 

To estimate the noise distribution of each data cube, and under the assumption of a Gaussian distribution for the noise, we fit a Gaussian (centered at $\rm S/N=0$) to the S/N histogram of the negative lines for the different width lines. To improve the statistics, we have grouped detections with different widths together before performing the fitting, specifically, we grouped lines with channels width $\rm N_{chn}$ 3 and 5, 7 and 9, etc. To better take into account the low number statistics in the tails of the noise distribution, we perform the fitting using a Poisson maximum likelihood estimator. 

We compute the reliability (or fidelity F) of a detection as:
\begin{equation}
\rm F (\rm S/N, N_{chn}) = 1 - \frac{N_{\rm neg}(\rm S/N, N_{chn})}{N_{\rm pos}(\rm S/N, N_{chn})}
\label{eq:fidelity}
\end{equation}
where $N_{\rm pos}(\rm S/N, N_{chn})$ and $N_{\rm neg}(\rm S/N, N_{chn})$ are the number of positive emission lines and the expected number of negative emission lines respectively, in each $\rm S/N$ bin. The expected number of negative emission lines is computed using a Gaussian function with the best fitted parameters found by the fitting process described above.

In some previous works \citep[e.g.][]{Walter2016,Loiacono2021}, the fidelity has been computed using the cumulative number counts (i.e.\,$N_{\rm neg}(\rm \geq S/N)$ and $N_{\rm pos}(\rm \geq S/N)$) instead of the differential number counts. However, if a cumulative approach is used, the computed fidelity at intermediate S/N values (for instance at $\rm S/N\sim 4.5$) would be completely dominated by the higher S/N detections (for instance a detection with $\rm S/N\sim 9.0$), resulting in a boost of the fidelity for lower S/N detections. To avoid this artificial boost of fidelity of intermediate S/N sources, here we use $N_{\rm neg}$ and $N_{\rm pos}$ as the differential number of detections in bins of S/N. 

\begin{figure}
\centering
\includegraphics[height=6.4cm]{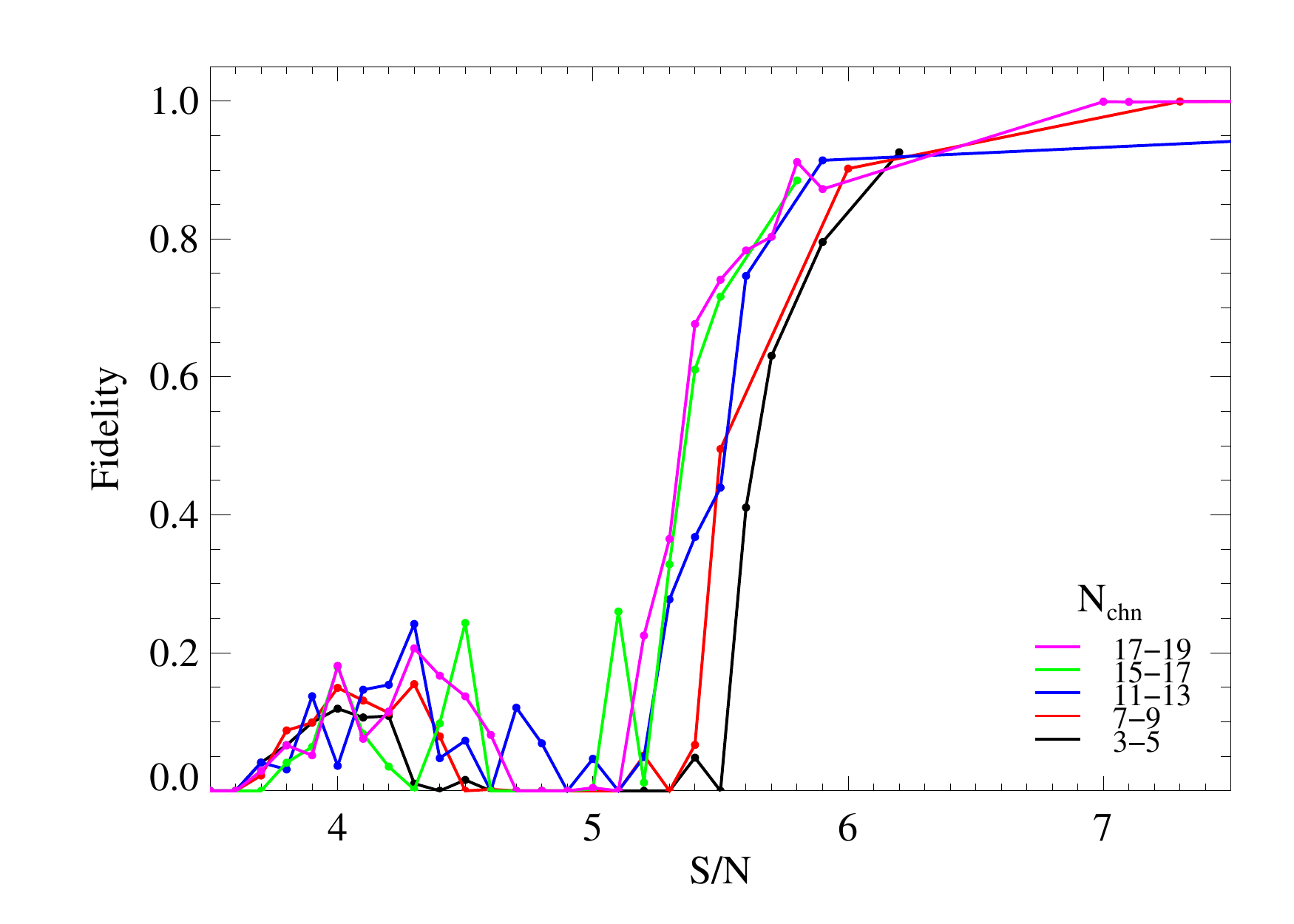} 
\caption{Median fidelity of a line candidate as a function of S/N for different line widths. In our analysis, we use a fidelity computed for each cube separately, but we show the median value in this figure for visualization. Taking into account all the fields together, we find that fidelity 0.8  is typically reached at $\rm S/N\sim5.6$, $\rm S/N\sim5.7$, $\rm S/N\sim5.8$, and $\rm S/N\sim6.0$ for sources with line widths $15-19$ ($\rm 371-470\,km\,s^{-1}$), $11-13$ ($\rm 272-322\,km\,s^{-1}$), $7-9$ ($\rm 173-223\,km\,s^{-1}$), and $3-5$ ($\rm 74-124\,km\,s^{-1}$) channels respectively. \label{fig:fidelity}}
\end{figure}

The noise properties varies among the 17 cubes, so we repeat this process for all our 17 cubes individually and determine the fidelity of each detected source in the cubes. For visualization purposes only, we computed the median fidelity over the cubes and show this in Fig.~\ref{fig:fidelity}. Since we compute fidelity using the differential number of detections, the fidelity cannot be constrained in S/N bins where no positive lines are detected. For computing the median, we only use the bins with constrained fidelity over the fields (data points in Fig.~\ref{fig:fidelity}) and we linearly interpolate between the S/N values.  

\subsection{Line Catalog}
\label{ssec:final_catalogs}
The choice of the fidelity threshold to keep a source in our catalogs is crucial and has to be taken carefully. This choice varies between different studies, depending on the aim of the analysis. \citet{Decarli2019} is aimed to constrain the CO luminosity function (used in our work as a reference for the computation of the background number counts as described in \S~\ref{ssec:cross}), and they choose a fidelity threshold of 0.2, but they treat the fidelity as upper limits and use a Montecarlo simulation to generate several realizations of their final catalog, in which they randomly assign a fidelity between 0.0 and their upper limits and only keep sources if this random fidelity value exceeds their threshold of 0.2. This means that lower fidelity sources are less likely to be kept in the catalog for each realization. 

This approach is adequate for measuring the luminosity function because even if sources with low fidelity (likely to be spurious) are included in some realizations, the overall statistics of the number counts will stay mostly unchanged (as explicitly checked in \citealt{Decarli2019}). Additionally, low fidelity sources are mostly composed of faint sources, only affecting the faint end of the luminosity function, which has a little impact on the overall fitting for the number counts. 

In the case of clustering measurements, secure detections are required, since all the sources contribute with the same weight to the final measurement, no matter their intrinsic flux. Including spurious detections would have a strong impact on the final measurement, especially for sparse samples. For this reason, we prefer to choose a more conservative fidelity threshold of $\geq$0.8, and we only keep sources fulfilling this criterion in our final catalogs. We nevertheless explore the impact of the fidelity threshold choice on our clustering results in Appendix \S~\ref{sec:impact}. 

From our median fidelity computation (see Fig.~\ref{fig:fidelity}), we find that fidelity 0.8  is typically reached at $\rm S/N\sim5.6$, $\rm S/N\sim5.7$, $\rm S/N\sim5.8$, and $\rm S/N\sim6.0$ for sources with line widths $15-19$ ($\rm 371-470\,km\,s^{-1}$), $11-13$ ($\rm 272-322\,km\,s^{-1}$), $7-9$ ($\rm 173-223\,km\,s^{-1}$), and $3-5$ ($\rm 74-124\,km\,s^{-1}$) channels respectively. 

Since the \textsc{findclumps} algorithm determines the source S/N using a global rms computed on the primary beam FWHM region of the ALMA pointing, the S/N of the detected candidates located close to the edges of the pointing may be overestimated due to noise fluctuation affecting these regions of the pointing. However, we still could detect reliable sources in these regions, so we include sources within the entire area covered by our observations, but we increase the fidelity threshold up to $\geq$0.9 for sources located in the edges of the pointing (where the telescope sensitivity is lower than 20\% of the maximum, or equivalently at radius larger than 50.14\arcsec\, from the central quasar). 

Considering all the fields together, our final catalogs contain 9 sources (excluding the detection of the central quasar). Note that this is the total number of sources detected in all the SPW0 cubes, but not all of them are necessarily included for the clustering computation (see details in \S~\ref{ssec:cross}). 

\begin{figure*}
\includegraphics[height=2.5cm]{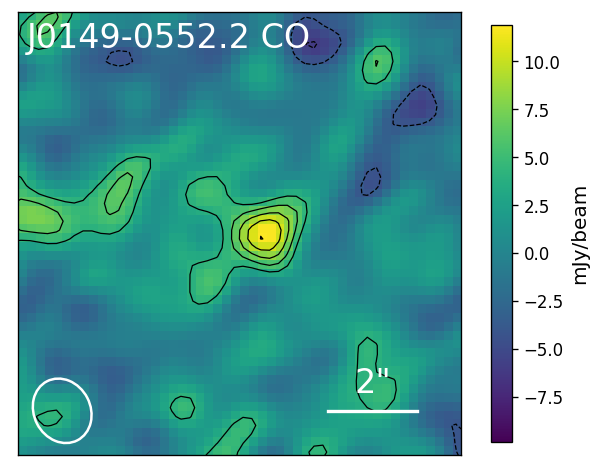}
\hspace{-0.4cm}
\includegraphics[height=3.3cm]{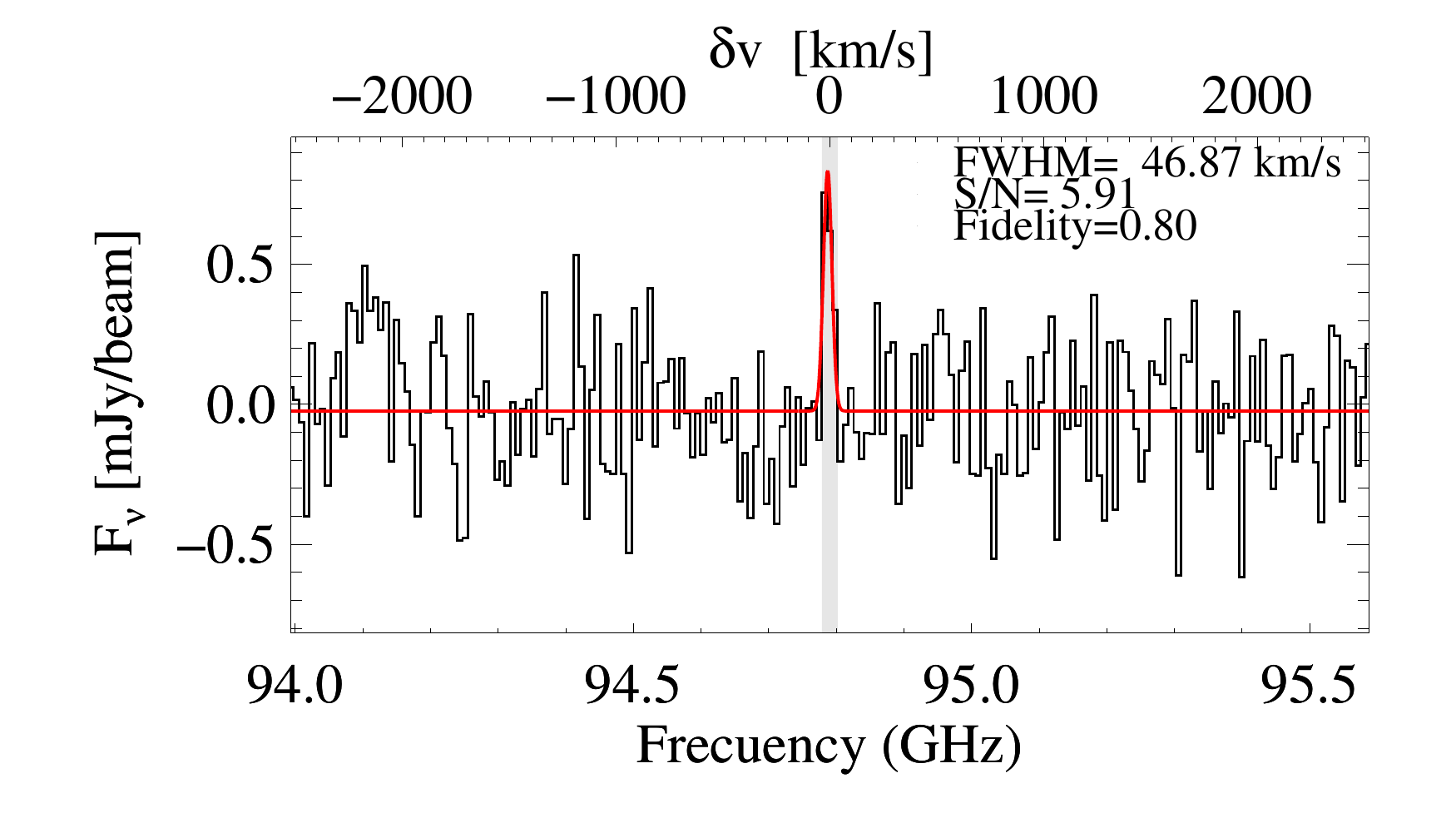}
\hspace{0.6cm}
\includegraphics[height=2.5cm]{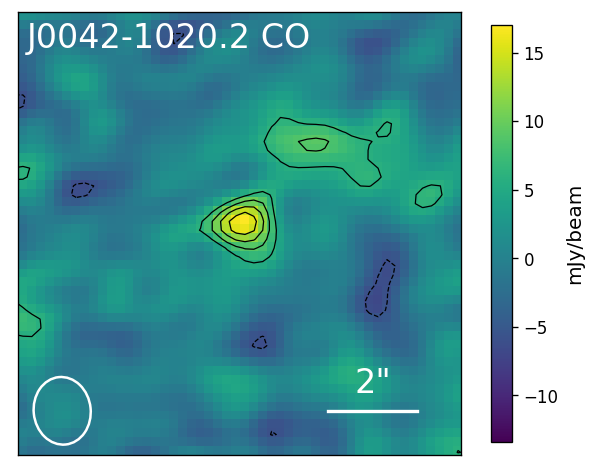}
\hspace{-0.4cm}
\includegraphics[height=3.3cm]{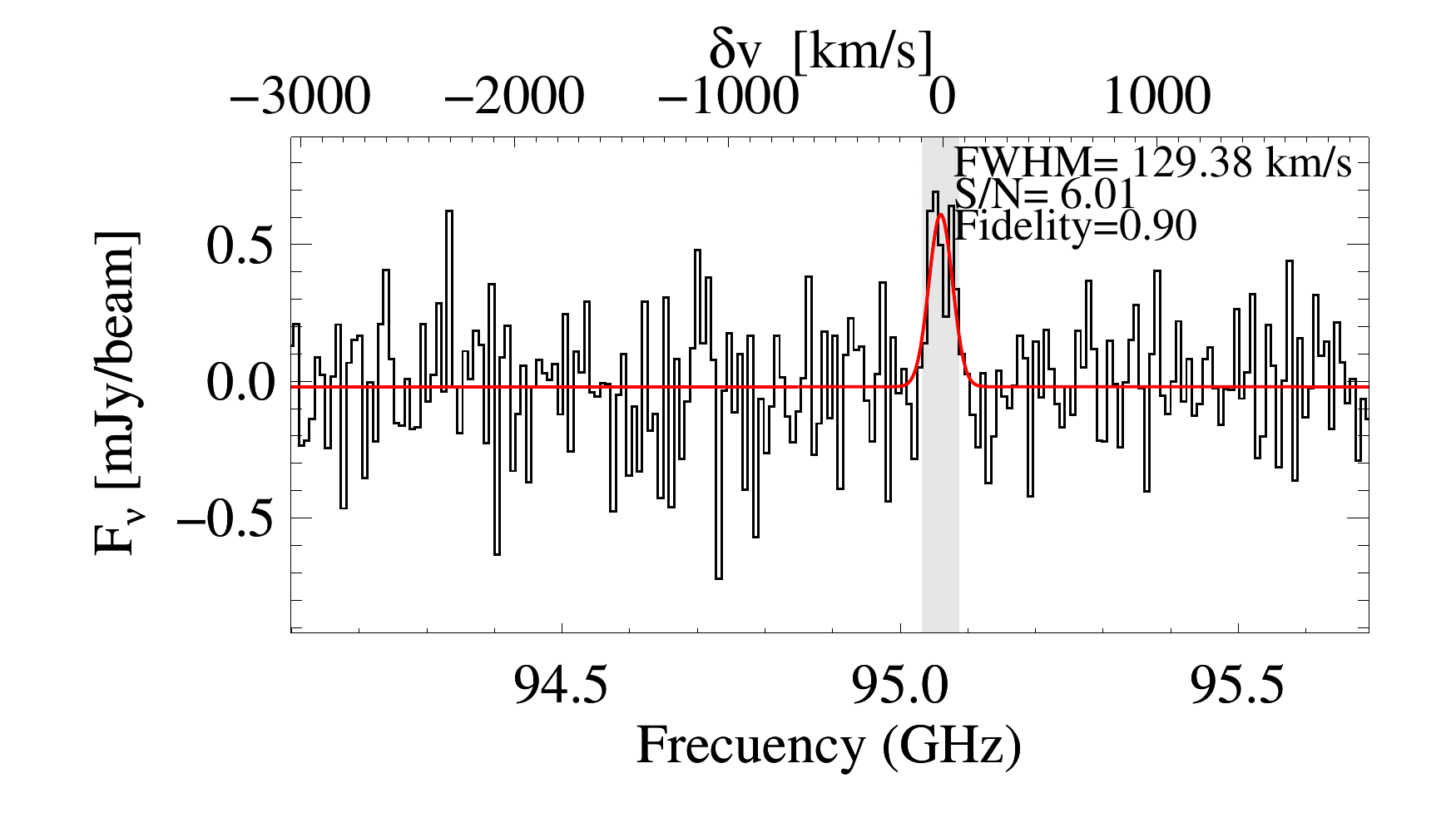}\\
\hspace{0.6cm}
\includegraphics[height=2.5cm]{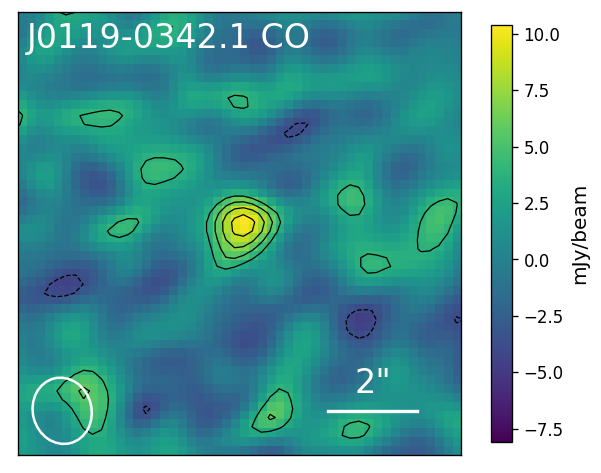}
\hspace{-0.4cm}
\includegraphics[height=3.3cm]{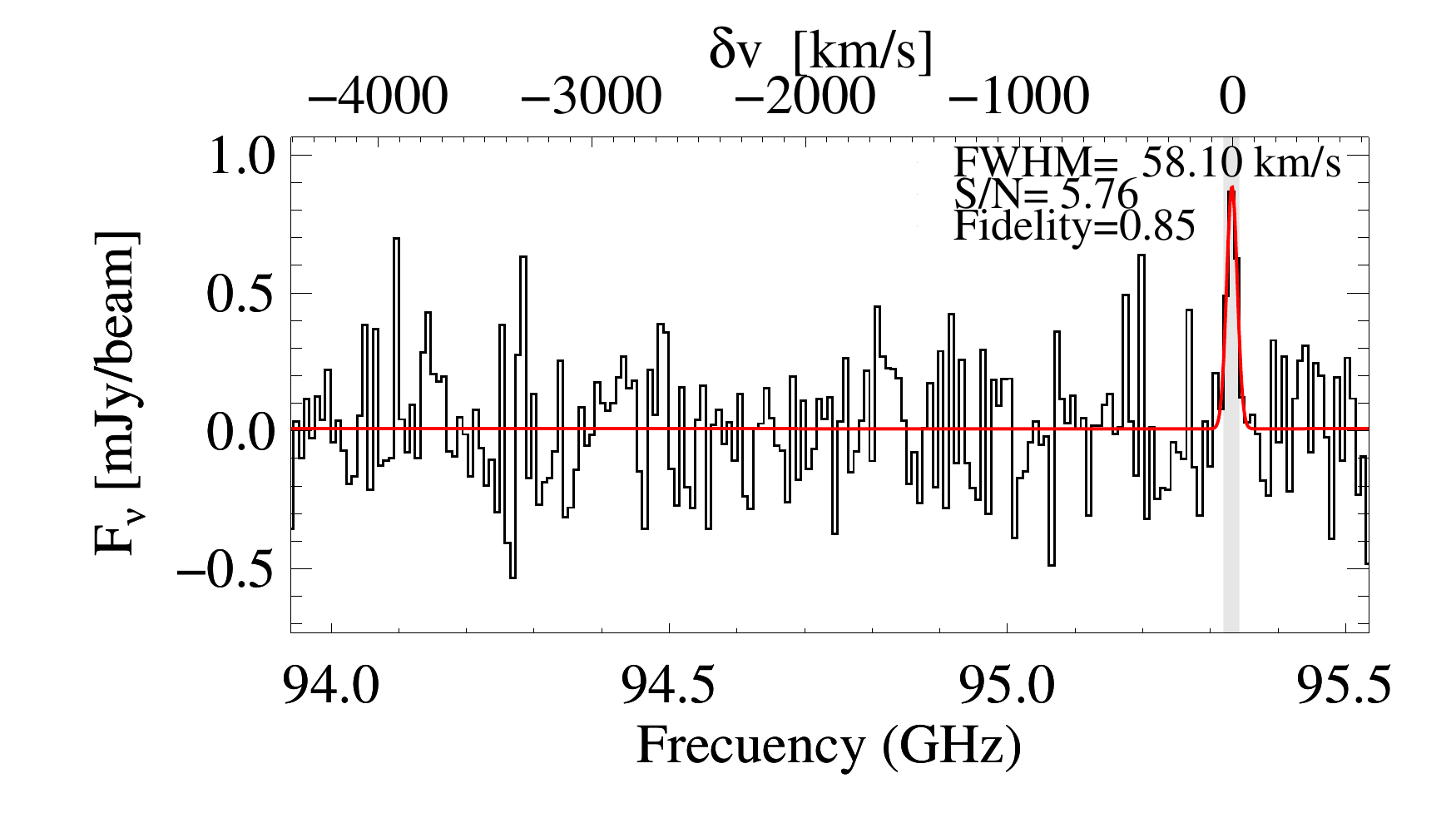}
\hspace{0.6cm}
\includegraphics[height=2.5cm]{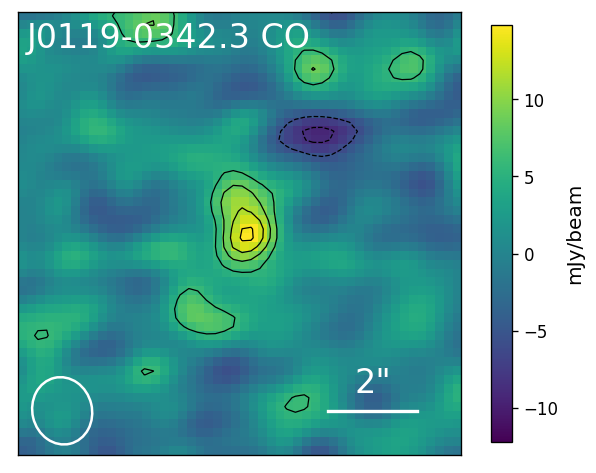}
\hspace{-0.4cm}
\includegraphics[height=3.3cm]{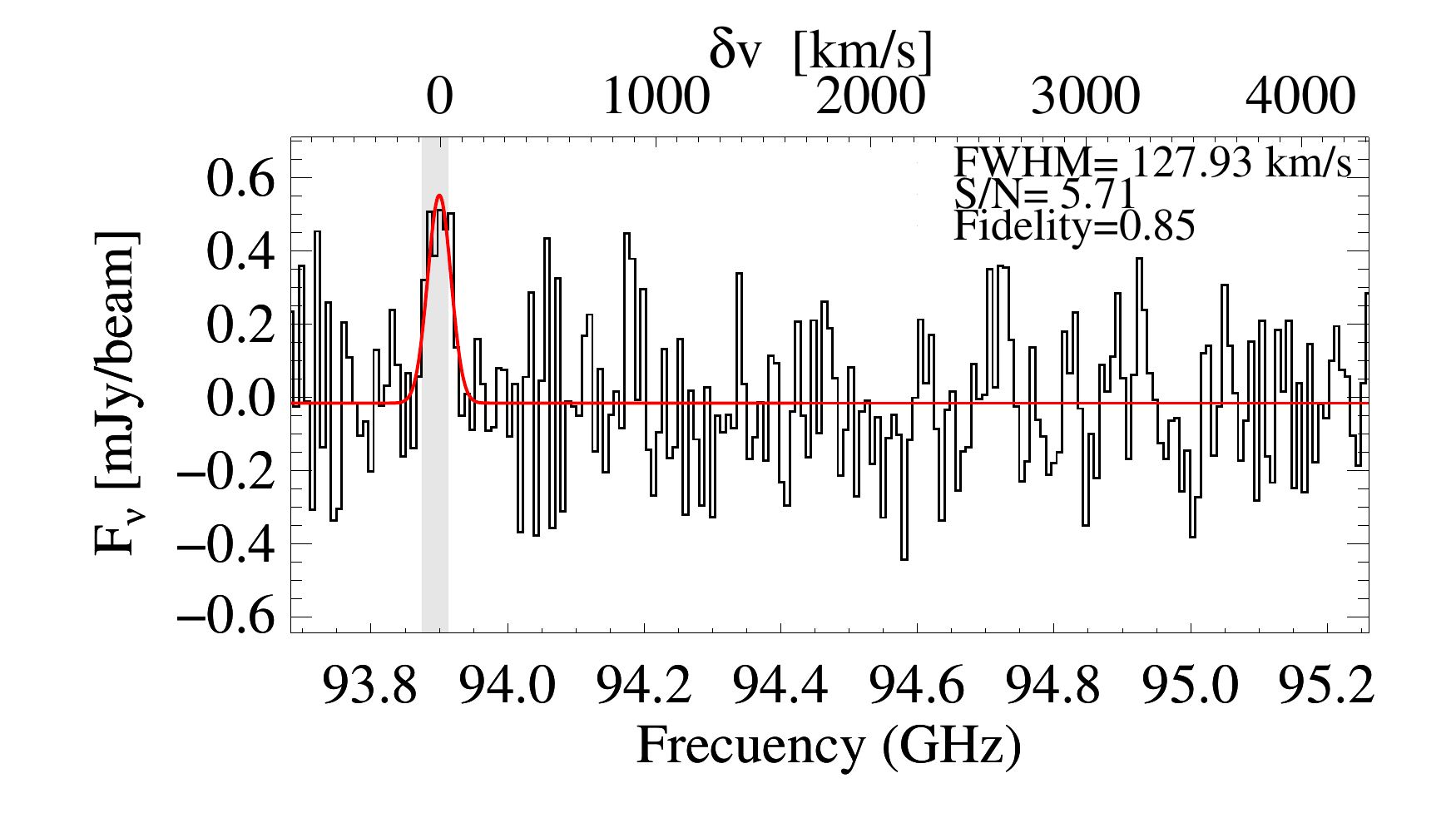}\\
\hspace{0.6cm}
\includegraphics[height=2.5cm]{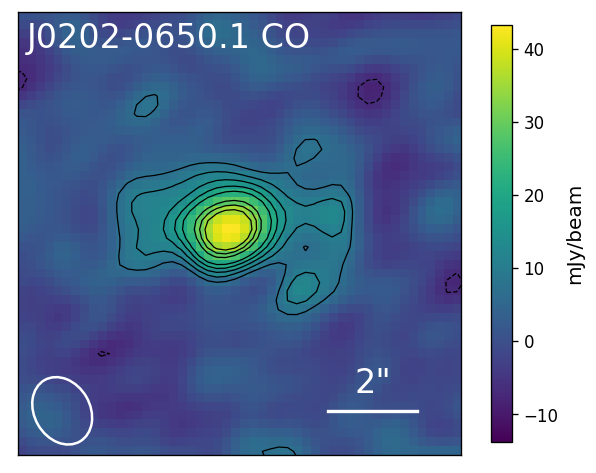}
\hspace{-0.4cm}
\includegraphics[height=3.3cm]{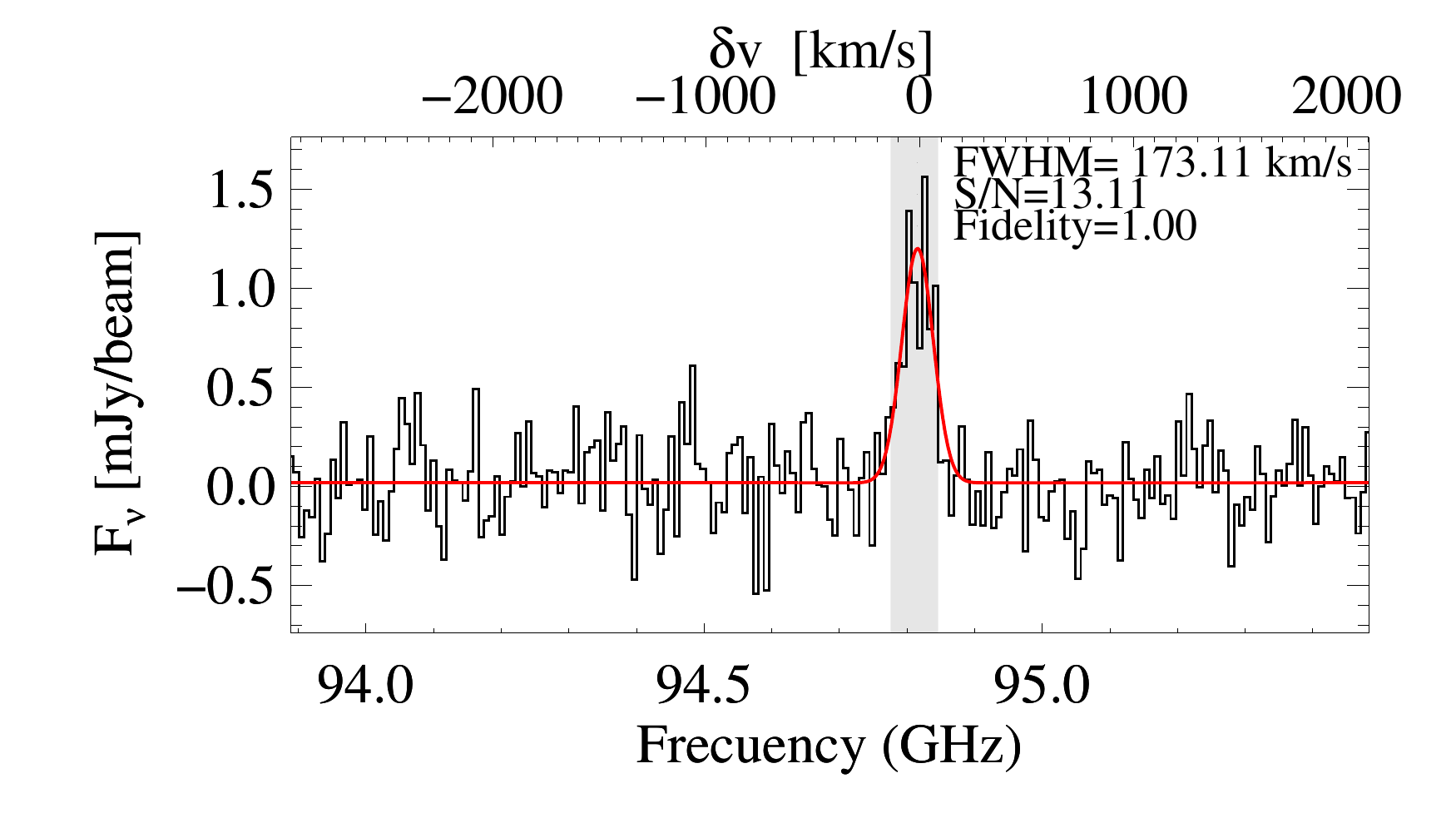}
\hspace{0.6cm}
\includegraphics[height=2.5cm]{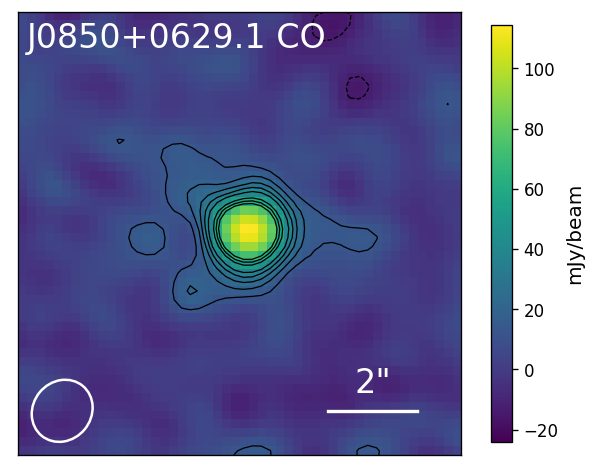}
\hspace{-0.4cm}
\includegraphics[height=3.3cm]{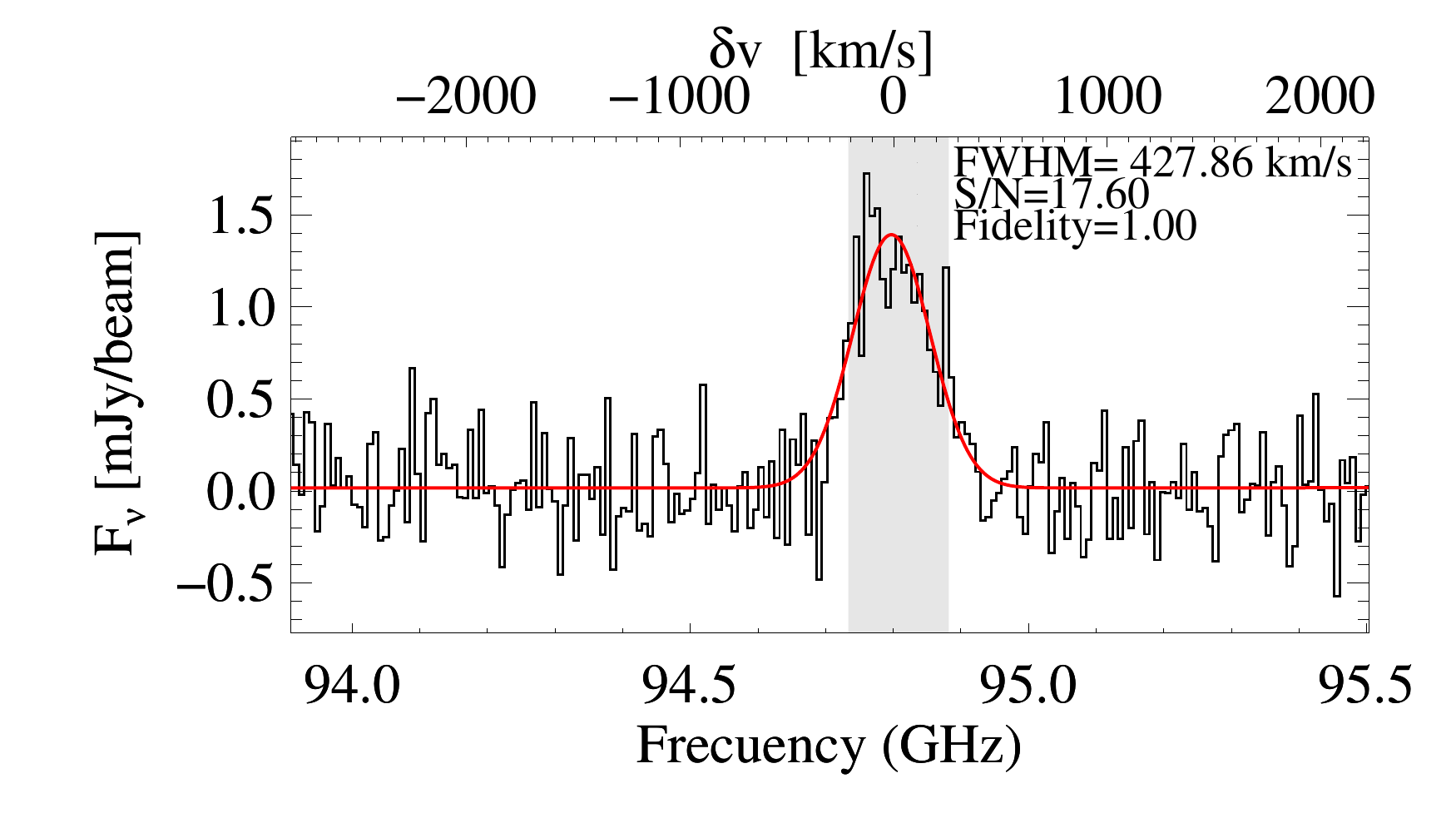}\\
\hspace{0.6cm}
\includegraphics[height=2.5cm]{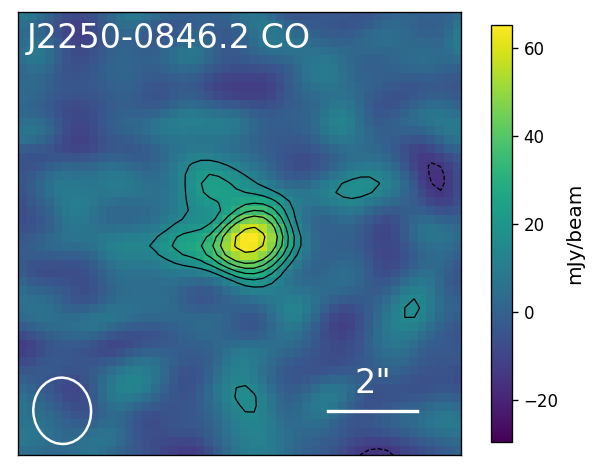}
\hspace{-0.4cm}
\includegraphics[height=3.3cm]{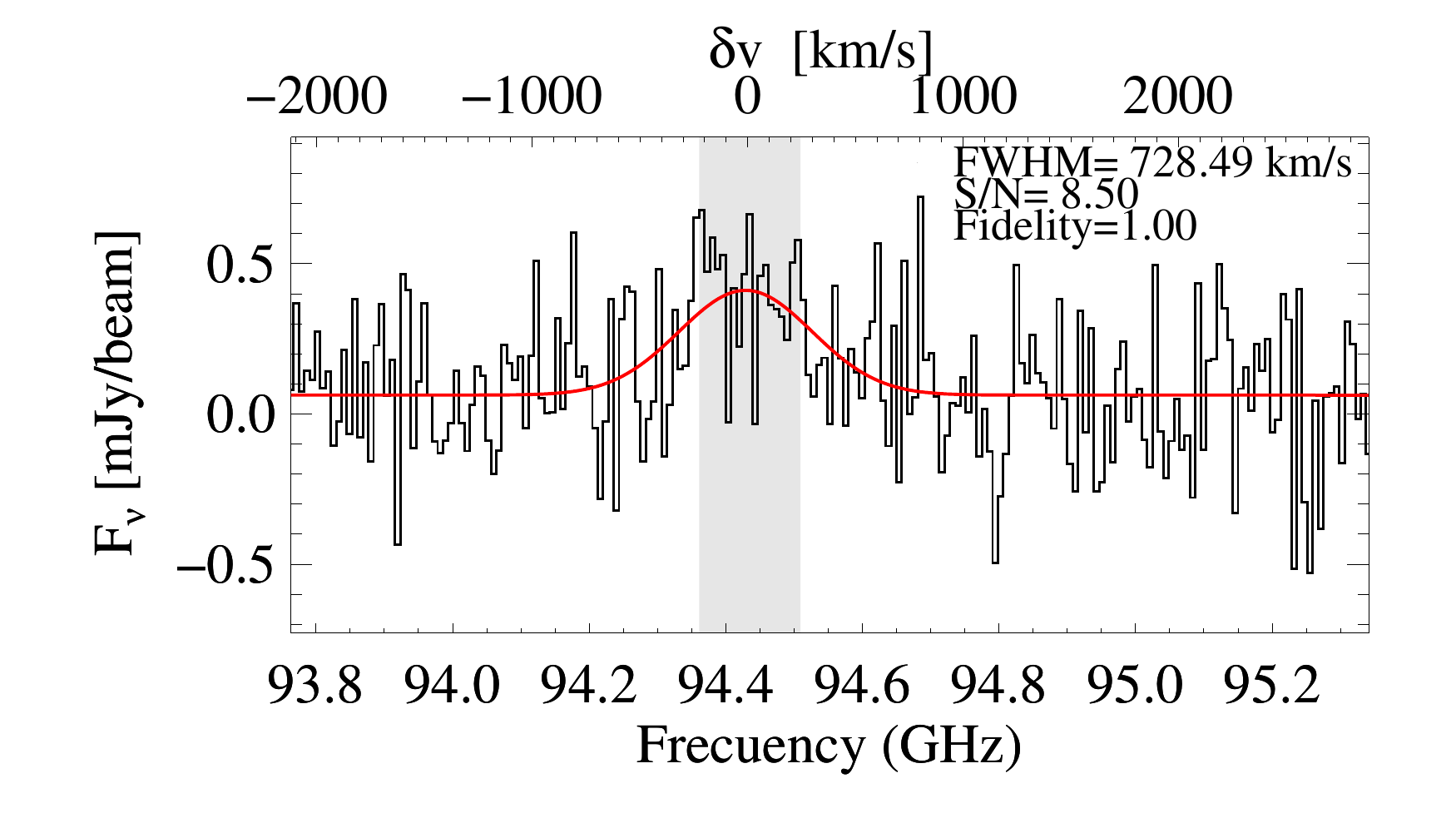}
\hspace{0.6cm}
\includegraphics[height=2.5cm]{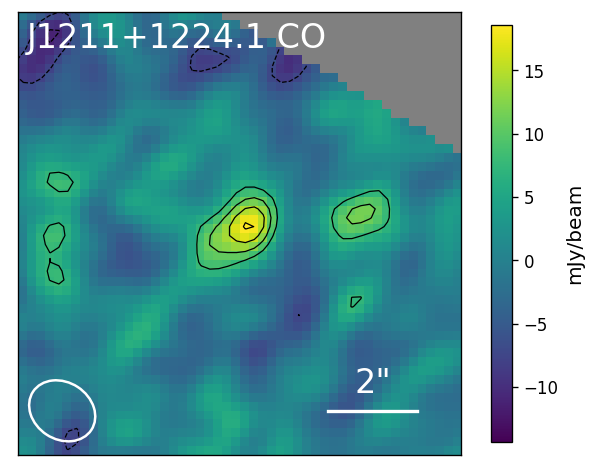}
\hspace{-0.4cm}
\includegraphics[height=3.3cm]{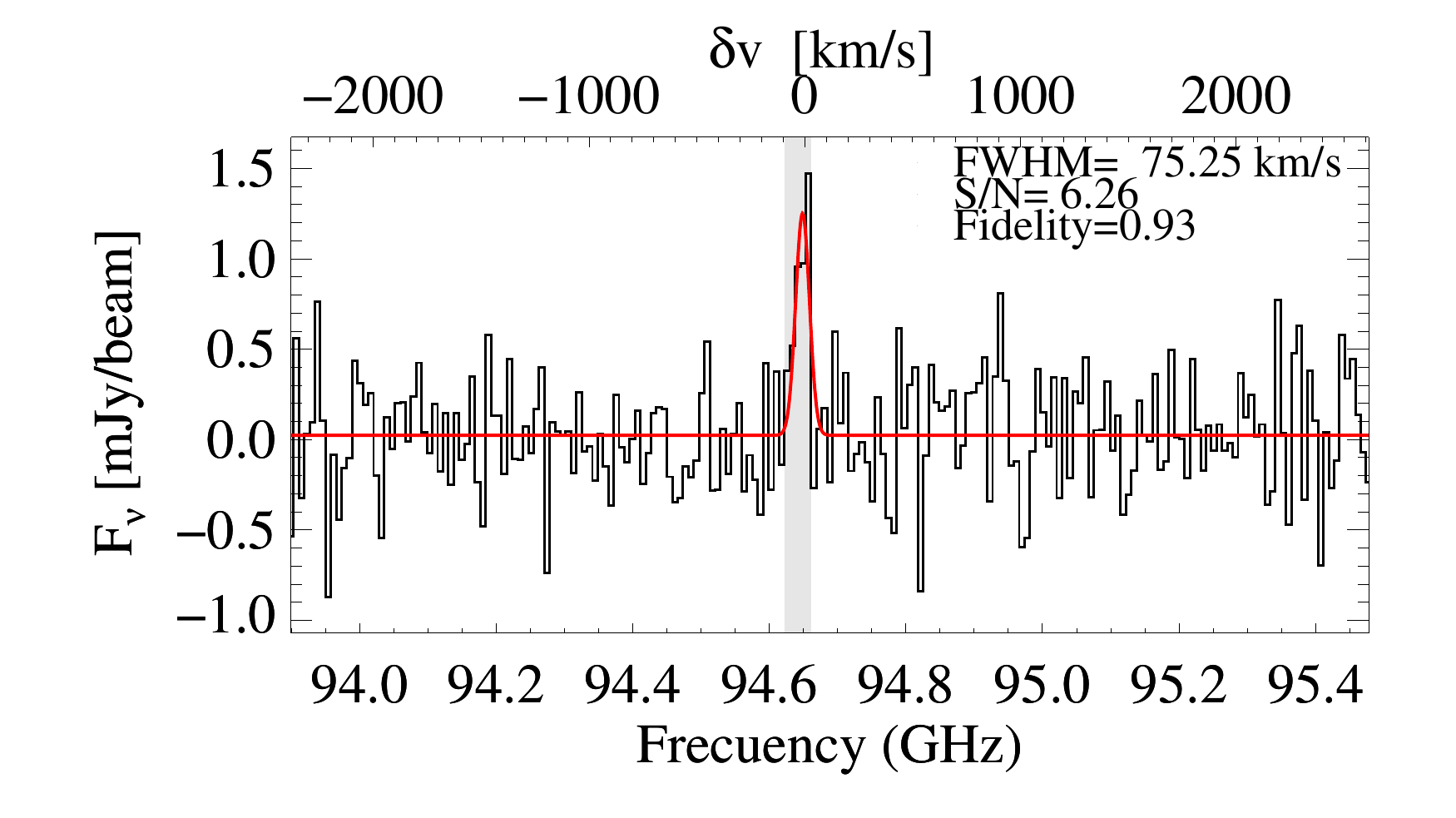}\\
\hspace{0.6cm}
\includegraphics[height=2.5cm]{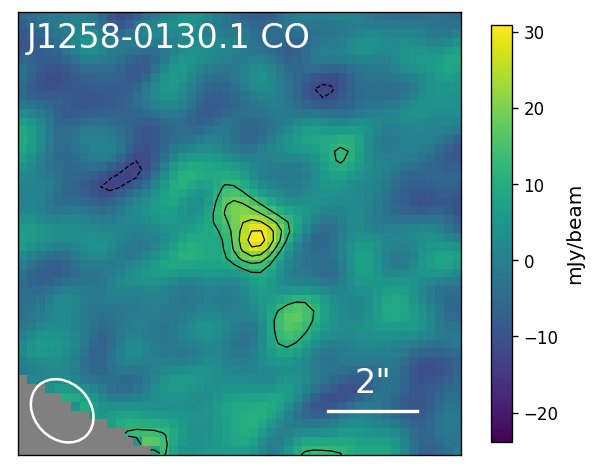}
\hspace{-0.4cm}
\includegraphics[height=3.3cm]{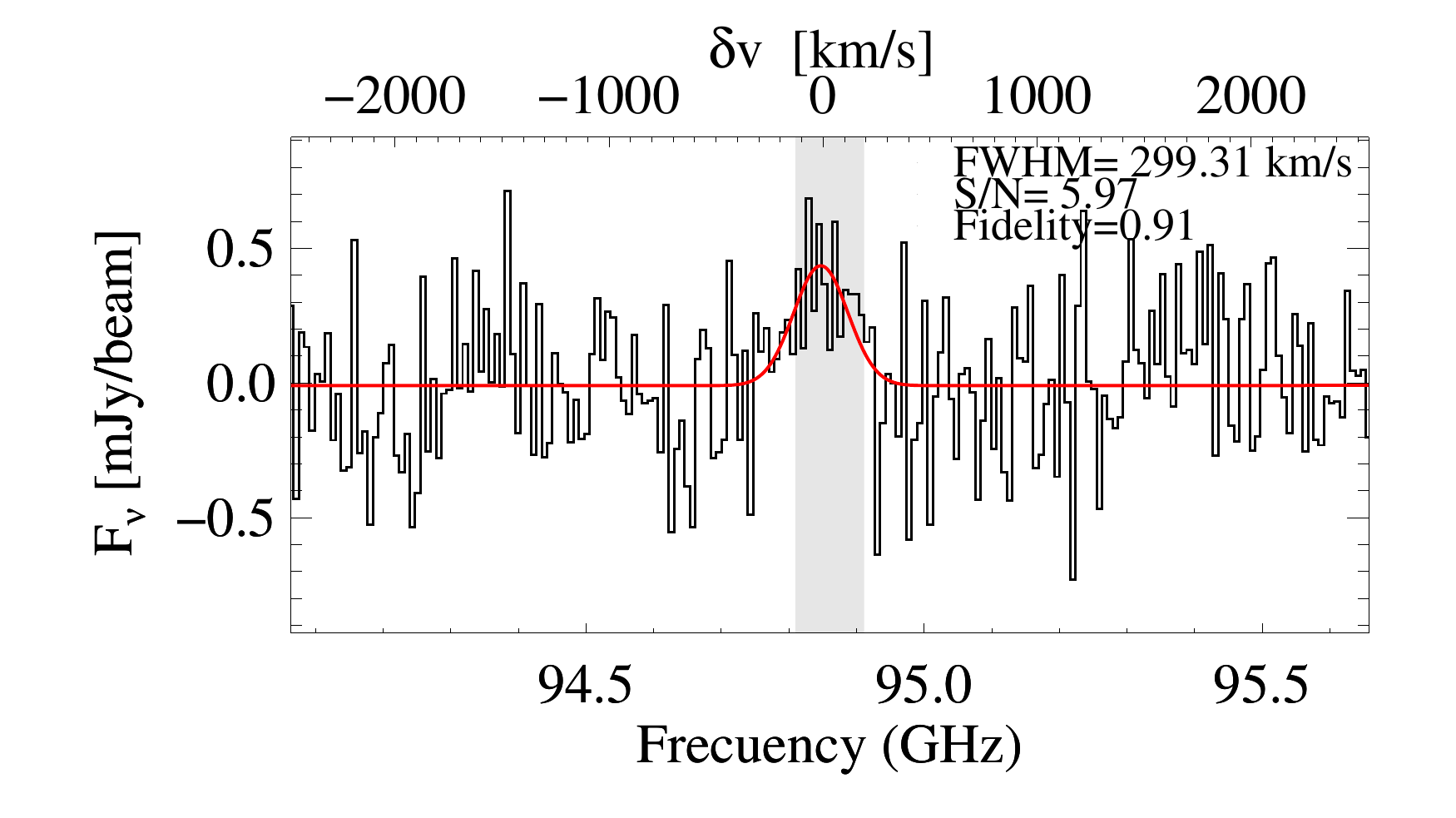}
\hspace{0.6cm}
\caption{\textit{Left:} Line maps for all the sources detected in our survey with fidelity $\geq$0.8 (or $\geq$0.9 if they are located at radius $>50.14$\arcsec from the pointing center), integrated over a frequency width given by $2.8\sigma_{\rm line}$ around the center of the line. Contours start at $\pm2\sigma$ and increase in steps of 1$\sigma$. The white ellipse in the lower-left corner shows the FWHM beam size. \textit{Right:} 1D extracted spectra on the brightest pixel of each source. We show the Gaussian plus flat continuum fit as a red curve. The shaded gray area shows the line width as detected by the line search algorithm. In each panel, we report the best-fitted FWHM value, the S/N of the line determined by the line search algorithm, and the fidelity. Line maps and spectra are both extracted from the cleaned cubes, and they are not corrected by the primary beam response. \label{fig:images_spec}}
\end{figure*}

We extract the 1D spectra for all the selected candidates from the clean cubes. At the resolution of our observations, the sources are not significantly resolved, therefore, we perform the spectral extraction on the brightest pixel of the source, and we perform a Gaussian fit with a flat continuum to constrain the line center ($\nu_{\rm line}$) and the line standard deviation ($\sigma_{\rm line}$). We use the \texttt{immoments} task from the \textsc{Casa} package to create a 0th moment map for each source. For that, the data cube is integrated over a frequency width given by $2.8\sigma_{\rm line}$ centered on $\nu_{\rm line}$ which recovers $\sim84\%$ of the line flux. In Fig.~\ref{fig:images_spec} we show the moment0 maps and the 1D extracted spectra for all the sources in our catalogs, and the best-fitted parameters for the lines are listed in Table~\ref{table:line_prop}. In Table ~\ref{table:n_per_field} we summarize the number of emission lines in each individual field. 

\begin{table*}
\caption{Properties of the emission lines detected in all the 17 fields with fidelity $\geq$0.8 (or $\geq$0.9 if they are located at radius $>50.14$\arcsec from the pointing center). \label{table:line_prop}}

\centering
\begin{tabular}{cccccccccc}
\hline
ID & RA (J2000) & Dec (J2000) & $\nu_{\rm line}$ & $\rm FWHM_{\rm line}$ & S/N & Fidelity &$\delta \theta$ & $\rm \delta v$ & Optical counterpart\\
& [deg] & [deg] & [GHz] & [$\rm km\,s^{-1}$] & & & [$\arcsec$] & [$\rm km\,s^{-1}$] &\\
(1)& (2) & (3) & (4) & (5) &(6) & (7)&(8) & (9) &(10)\\
\hline 
J0042-1020.2 & 00:42:18.48 & -10:20:36.48 &  95.06 & 129 &   6.01 & 0.90 &  33.0 &  -1767.19 & $R>25.6$, $g>26.3$\\
J0119-0342.1 & 01:20:00.37 & -03:42:59.90 &  95.33 &  58 &    5.76 & 0.85 &  45.1 &  -2266.36  & $R>25.7$, $g>26.4$\\
J0119-0342.3 & 01:20:02.15 & -03:42:16.96 &  93.90 & 128 &   5.71 & 0.85 &  39.0 &   2270.94  & $R>25.7$, $g>26.4$\\
J0149-0552.2$^{*}$ & 01:49:08.41 & -05:52:41.34 &  94.79 &  47 &    5.91 & 0.80 &  31.3 &   -851.97  &$R>25.7$, $g>26.5$\\
J0202-0650.1$^{*}$ & 02:02:54.72 & -06:50:51.88 &  94.82 & 173 &  13.11 & 1.00 &  16.1 &   -829.85 & $R=24.4$, $g=25.7$\\
&&&&&&&&& Lyman break detection \\
&&&&&&&&&with $g-R=1.3$ located at 1.2\arcsec\\
J0850+0629.1$^{*}$ & 08:50:14.22 & 06:29:49.93 &  94.80 & 428 &  17.60 & 1.00 &  11.9 &   -710.84&$R>25.7$, $g>26.3$\\
J1211+1224.1$^{*}$ & 12:11:45.06 & 12:25:06.25 &  94.65 &  75 &    6.26 & 0.93 &  54.9 &    559.31  & $R>25.3$, $g>26.3$\\
&&&&&&&&& LAE located at 1.6\arcsec\\
J1258-0130.1 & 12:58:43.81 & -01:31:09.35 &  94.85 & 299 &   5.97 & 0.91 &  53.1 &  -1220.26 & $R>25.7$, $g>26.4$\\
J2250-0846.2$^{*}$ & 22:50:50.92 & -08:45:40.39 &  94.43 & 728 &   8.50 & 1.00 &  32.7 &    398.63 &$R>25.6$, $g>26.4$\\
\hline
\multicolumn{4}{l}{$^{*}$ included for the clustering measurement.}
\end{tabular}
\tablecomments{Columns (1), (2), and (3) indicate the name of the candidate and the RA, Dec sexagesimal coordinates. Columns (4) and (5) indicate the central frequency and the FWHM of the detected line as determined by the best Gaussian+continuum fitting. Column (6) shows the S/N of the detection as determined by the line search algorithm. Column (7) shows the fidelity computes using eqn.~(\ref{eq:fidelity}). Column (8) and (9) show the angular distance and the velocity distance respectively between the emission line and the central quasar, assuming that the observed line corresponds to the CO(4--3) transition. For the computation of the velocity distance, we use the redshift based on the ALMA detection of the quasar, when it is detected with fidelity $\geq0.8$, otherwise, we use the redshift as determined from the optical quasar spectrum. Column (10) shows the optical color magnitude $g$ and $R$ measured in a $2\arcsec$ aperture in the position of the source (or at the position of a counterpart if one exists within $2\arcsec$). The lower limits for the magnitude correspond to $3\sigma$ limits. We provide comments about optical counterparts for the source.}
\end{table*}

One of the sources (J2250-0846.2) is identified by our algorithm as two close ($\delta \theta=0.4\arcsec$, $\rm \delta v\sim 625\rm\,km\,s^{-1}$) sources. This could be either a galaxy merger or a galaxy with a two peaked emission line, with total width FWHM$\rm >700 \,km\,s^{-1}$. In this study, we consider this as a single object, since even in the merger scenario this would likely trace the same dark matter halo. We study this object further in a forthcoming paper (Garc\'ia-Vergara et al. in preparation).

Our candidates span a large range of line widths, ranging from $\rm 47\,km\,s^{-1}$ to $\rm 728\,km\,s^{-1}$, with three of the lines showing small FWHM values ($\rm <100\,km\,s^{-1}$)\footnote{We note that one of these narrow lines is the object J1211+1224.1, which has a LAE counterpart at the same redshift.}. To explore if the existence of such narrow lines is expected and reasonable, we checked the line widths of the 16 CO emitting galaxies blindly found in the ASPECS survey with fidelity $>0.9$ \citep{Gonzalez-lopez2019}. We find that their FWHM range from $\rm 40\,km\,s^{-1}$ to $\rm 617\,km\,s^{-1}$, including two very narrow lines with $\rm 40\,km\,s^{-1}$ (with $\rm S/N=7.9)$ and $\rm 50\,km\,s^{-1}$ (with $\rm S/N=9.5)$ respectively. Using HST counterparts for the CO emitters detected in ASPECS, \citet{Aravena2019} suggest that these galaxies are likely very face-on, which would explain the detection of such narrow line widths in CO emitting galaxy surveys. 

\subsection{Line Identification and Contamination}
\label{ssec:contamination}

As we targeted quasars with known redshifts, which are expected to be surrounded by large overdensities of galaxies, and given that we are tracing relatively small areas in the sky, we assume that all the detected lines correspond to the CO(4--3) transition and that the serendipitous detection of emission lines at other redshifts is negligible. To support this assumption, here we quantify the probability that each of our detected lines is actually a galaxy at a different redshift, and we also use our optical images to look for a counterpart of the detected emission lines.

\subsubsection{Probability of Low-z Contaminants}
\label{sssec:cont}
First, we compute the probability that each source in our catalog is a low-z galaxy. For this computation, we first explore what transitions at low-z could be detectable given the frequency setup of our survey, and we compute the comoving volume traced for them in each quasar field. We consider a circular area given by $R\leq59\arcsec$ (the same used to look for emission lines) and a redshift coverage given by i) $\rm dv=\pm3000\rm\,km\,s^{-1}$, corresponding to the width of the whole cube, the same where we detect our 9 candidates, and ii) $\rm dv= \pm1000\rm\,km\,s^{-1}$, corresponding to the width used for the clustering analysis, where we detect 5 candidates (see details in \S~\ref{ssec:cross}). We summarize this information in Table~\ref{table:contam}. The main possible contaminants in our sample are the CO transitions CO(1--0) at $z\sim0.2$, CO(2--1) at $z\sim1.4$, and CO(3--2) at $z\sim2.7$.

To compute the probability, we use the CO luminosity function of the different CO transitions \citep{Decarli2019} at the corresponding redshift, and integrate it down to the $5.6\sigma$ limiting luminosity of our survey (this is the same S/N threshold used in this work for the detections of CO emitters, see \S~\ref{ssec:final_catalogs}) to compute the number density of sources for each transition. We multiply this quantity by the traced comoving volume to obtain the expected number of contaminants in one field, and we show the result in Table~\ref{table:contam}. The limiting luminosity at the center of the pointing is reported in Table~\ref{tab:obs}. Note that the limiting luminosity varies over the ALMA pointing due to the decrease of sensitivity with the radius, thus we include this in our computation, following the procedure described in \S~\ref{ssec:cross}. 

The probability that each of the lines presented in our final catalog is CO(1--0), CO(2--1), and CO(3--2) is $2$\%, $25$\%, and $10$\% respectively. If we only focus on the sample used for the clustering analysis composed of 5 sources, we find that the probability that each of these sources is CO(1--0), CO(2--1), and CO(3--2) is $0.8$\%, $8.3$\%, and $3.4$\% respectively. 
 
\begin{table}
\caption{Probability of the presence of low-z contaminants in our survey.  \label{table:contam}}
\centering
\begin{tabular}{ccccccc}
\hline 
Transition & $z_{\rm min}$ & $z_{\rm max}$ & Volume$_{3000}$ & $N_{3000}$ & $N_{1000}$ \\
  &   &  &  [$h^{-1}\rm cMpc$ ] &   &   \\
 (1)   &  (2)   &  (3) &   (4)&  (5)  &  (6)  \\
\hline
CO(1--0) & 0.206 & 0.231 & 6.486      & 0.023 & 0.008\\
CO(2--1) & 1.413 & 1.461 & 147.276  &0.250 & 0.083\\
CO(3--2) & 2.619 & 2.691 & 253.083  &0.103 & 0.034\\
\hline
\end{tabular}
\tablecomments{For the different transitions, we report the redshift and the comoving volume traced in one quasar field at the redshift of the transition. The comoving volume is given by a circular area with radius $R\leq59\arcsec$ and a redshift coverage given by $\rm dv=\pm3000\rm\,km\,s^{-1}$. Column (5) and (6) show the expected number of lines per quasar field in the comoving volume indicated in column (4), and in 1/3 of that volume (i.e.\,assuming $\rm dv= \pm1000\rm\,km\,s^{-1}$) respectively. Note that in absence of clustering the number of expected lines only depends on the volume, therefore the number of lines simply decrease by a factor of 3 in column (6) compared with column (5).}
\end{table}

\subsubsection{Optical Counterparts}
\label{sssec:opt}
The second approach that we use to explore possible contamination from low-z galaxies is to cross-match the spatial position of the detected lines with the position of LAEs detected in the optical images available for our 17 fields. We limit the search for the optical counterparts of our emission lines to a radius of 2\arcsec. Using the available LAE catalog (see section \S~\ref{ssec:obs_opt}), we find that one of the detected candidates (J1211+1224.1) coincides with the position of an LAE detected in our previous survey.  We also explored the optical images at the position of all the line detections, but we do not find any LAE candidate at these positions, even when relaxing the LAEs detection down to $\rm S/N>3$. 

We also check the possibility to include additional low fidelity line emission sources at the position of the only other $\rm S/N>5$ LAE that were located within the ALMA Field-of-view. However, even by relaxing the fidelity criteria down to 0.3 in the ALMA catalog, no line detections were found within $3\arcsec$ from the position of the LAE. 

Since the UV-continuum emission of $z\sim4$ galaxies is characterized by a strong flux break at $\lambda_{\rm RF}=1216\AA$ (the so-called Lyman break) owing to the absorption of photons with $\lambda_{\rm RF}\leq1216\AA$ by neutral hydrogen in the intergalactic medium, we can use the $R$ and $g$ band optical images to detect the expected flux break. Although all the galaxies should exhibit this break, not all of them will exhibit the LAE emission line, and thus may not be included in the LAE catalog. Note that the presence of the flux break by itself can not confirm the redshift of the source, since we would need an additional observation on the UV-continuum at longer wavelengths to exclude possible interlopers, but the absence of the break can be used to confirm that the source is located at lower redshift. 

For all the detected lines we search for a $\geq3\sigma$ detection in the $R$ and $g$ images within $2\arcsec$. If a source is found in either of these bands, we perform a $2\arcsec$ aperture photometry on that position in the two bands following the same criteria as in \citet{Garcia-Vergara2019}. If no source is found, we perform the aperture photometry at the position of the detected line and compute $3\sigma$ upper limits for the fluxes in the $g$ and $R$ bands. In all the cases, we compute the color magnitude $g-R$ and report these values in Table~\ref{table:line_prop}. As in \citet{Garcia-Vergara2019}, the Lyman break is defined by the color criteria $g-R\geq0.7$. 

In addition to the LAE mentioned above (J1211+1224.1), we find that one other CO emitting galaxy (J0202-0650.1) exhibits the Lyman break, suggesting that it is possibly an LBG at $z\sim4$. All the other galaxies are not detected in either $g$ and $R$ bands, so we do not have information to trace the Lyman break, and we can not confirm that any of these are low-z interlopers. However, the lack of an optical counterpart at the VLT survey (with $3\sigma$ limiting magnitude 25.61 and 26.37 for the $R$ and $g$ images respectively) makes these objects unlikely to be lower$-z$ ($z\sim0.2$ for CO(1--0), $z\sim1.4$ for CO(2--1), and $z\sim2.6$ for CO(3--2); see Table~\ref{table:contam}) galaxies. 

Specifically, we check the typical $R$ magnitude for  CO(2--1) emitters at $z\sim1.4$, which are our main potential contaminants (see Table~\ref{table:contam}). We used the photometric catalog available from \citet{Skelton2014} to check the median $R$ magnitudes\footnote{$R$ magnitudes correspond to that measured in images taken using ESO/WFI with the filter $R_c$ (see \citealt{Skelton2014} for details).} for the 11 CO(2--1) sources detected with high fidelity in the ASPECS survey \citep[see Table~6 in][]{Gonzalez-lopez2019}. We find that these sources have a median magnitude $R=24.83$ which is much lower than the $3\sigma$ limiting magnitude in the R band for our candidates (see Table~\ref{table:line_prop}), making them unlikely to be CO(2--1) emitters at $z\sim1.4$.

\subsection{Quasar Detection}
We detect an emission line in 10/17 quasars, which we assume to be the CO(4--3) transition as they all have an optical counterpart and spectroscopic redshift from SDSS spectra. All 10 lines have fidelity $\geq$0.8 and are detected with S/N ranging from 5.9 up to 24.5. We relax our S/N criteria down to $\rm S/N>4$, and we look for possible line candidates for the other 7 quasars, restricting the search to detections with line widths $\rm \geq 124\,km\,s^{-1}$. When more than one candidate was found at the expected sky projected position of their optical counterpart (within $3\arcsec$), we selected the one with higher S/N. We do not impose a requirement in the velocity space position of the quasar counterpart, since the redshift of optical quasars could be associated with larger uncertainties than reported in previous studies. An uncertainty in the optical computed redshift of $\rm 2970 \,km\,s^{-1} $ (one half of the bandwidth) could be still possible. This search results in the detection of all the other 7 quasar counterparts, detected with S/N ranging from 4.1 up to 5.3, but with low fidelity ($<0.4$), and therefore some of these detections could be noise fluctuations. 

\begin{figure}
\centering
\includegraphics[height=6.5cm]{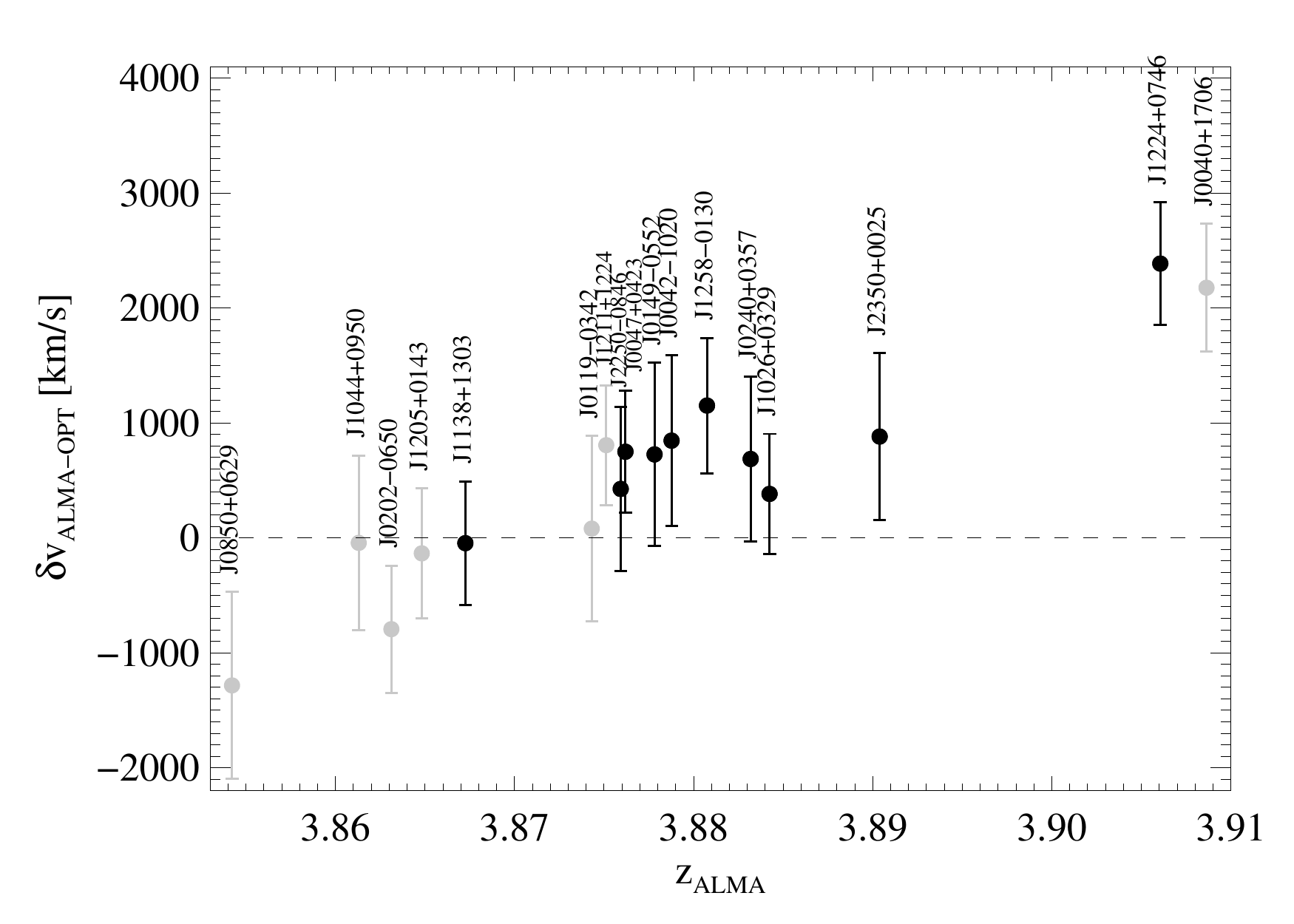} 
\caption{Velocity offset between the quasar redshifts determined from the CO(4--3) emission lines and the rest-frame UV emission lines. We include the secure quasar detections (black) and the low-fidelity detections (gray). Based only on the secure detections, we find a median offset of $|\rm \delta v|=738 \pm 651 \rm\,km\,s^{-1}$.\label{fig:vel_offset_qso}}
\end{figure}

We compute the offset between the CO(4--3) redshift and the rest-frame UV emission lines based redshift for all the quasars and we show our results in Fig.~\ref{fig:vel_offset_qso} and Table~\ref{table:n_per_field}. For the secure quasar detections, we find a median velocity offset of $|\rm \delta v|=738 \pm 651 \rm\,km\,s^{-1}$. This is consistent with the typical reported uncertainties of the optical-based redshifts (see Table~\ref{table:quasar}). The only exception is the quasar located in the field J1224+0746 which exhibits a large offset of $|\rm \delta v|=2386\rm\,km\,s^{-1}$, and therefore is located at $\rm 1004\,km\,s^{-1}$ from the edge of the SPW0. For completeness, we also include the velocity offset computed for the quasar detected with low fidelity in our ALMA observations, but in this study, we only use the redshifts from the secure detections. We present a detailed analysis of the quasar properties in a forthcoming study (Garc\'ia-Vergara et al. in preparation). 

\begin{table*}
\caption{Quasar redshift determined from ALMA observations, and number of emission lines in each field. \label{table:n_per_field}}
\centering
\begin{tabular}{ccccccc}
\hline 
Field & $z_{\rm opt}$ & $z_{\rm ALMA}$& $\rm \delta v\,[\rm km\,s^{-1}]$ & $N_{3000\,} (\delta_{\rm CO})$ & $N_{1000}\, (\delta_{\rm CO})$ & $\delta_{\rm LAE}$\\
(1) & (2)  & (3) &(4)  & (5)  & (6)  & (7) \\
\hline
J0040+1706 &  3.873 &   3.910$^*$           &  2266$^*$            & 0\, (00.00)   &0\, (00.00) &2.66\\
J0042-1020 &  3.865 &  3.878   						  &    805     					& 1\, (17.77)  &0\, (00.00) &0.00\\
J0047+0423 & 3.864 &  3.877    					   &   780      				& 0\, (00.00) &0\, (00.00) &0.00 \\
J0119-0342 &  3.873 & 3.874$^*$             &   46$^*$           & 2\, (34.90)  &0\, (00.00) &3.41\\
J0149-0552 &  3.866 &   3.879   					   &  781      					& 1\, (18.61)  &1\, (55.87) &0.00\\
J0202-0650 &  3.876 &  3.864$^*$           &   -741$^*$         & 1\, (17.63)    &1\, (52.91) &0.00 \\
J0240+0357 &  3.872&  3.883    					 &   661      					& 0\, (00.00)  &0\, (00.00) &0.92 \\
J0850+0629 &  3.875&  3.854$^*$            &  -1310$^*$    & 1\, (19.40)    &1\, (58.22) &2.41\\
J1026+0329 &  3.878&  3.884   					      &    380     			    & 0\, (00.00)  & 0\, (00.00) &2.16\\
J1044+0950 &  3.862&  3.861$^*$           &   -37$^*$       & 0\, (00.00)   &0\, (00.00) &0.78 \\
J1138+1303 &  3.868&  3.867    					 &   -62       			    & 0\, (00.00) &0\, (00.00) &0.00 \\
J1205+0143 &  3.867&  3.864$^*$            &  -161$^*$     & 0\, (00.00)  &0\, (00.00) &1.34 \\
J1211+1224 &  3.862&   3.875$^*$          & 805$^*$         & 1\, (25.97)  &1\, (77.95) & 3.75\\
J1224+0746 &  3.867&   3.905    					&   2355    				& 0\, (00.00) &0\, (00.00) & 0.00\\
J1258-0130 &  3.862 &   3.882   					&   1203     				& 1\, (27.34) &0\, (00.00) & 1.08\\
J2250-0846 &  3.869 &   3.876   					&   433      				& 1\, (19.61) &1\, (58.86) & 0.77\\
J2350+0025 & 3.876 &   3.891   					&    920      				& 0\, (00.00) &0\, (00.00) & 0.59\\
ALL               &            &              					 &                				&9\, (10.56) &5\, (17.60)  &1.36\\
\hline
\multicolumn{7}{l}{$^{*}$ Indicates the cases where the CO(4--3) line is detected with low ($<0.8$) fidelity.}
\end{tabular}
\tablecomments{For each field, column (2) shows the quasar redshift determined from UV rest-frame emission lines, column (3) show the redshift determined from the CO(4--3) emission line, column (4) shows the velocity offset between these (with typical uncertainties of $\rm 651\,\rm km\,s^{-1}$) , column (5) and (6) show the number of CO emitters and the corresponding overdensity within $\pm3000\rm\, km\,s^{-1}$ and $\pm1000\,\rm km\,s^{-1}$ from the quasar redshift, respectively (using the redshift determined from the CO(4--3) line if detected with high fidelity ($\geq0.8$) and the UV rest-frame emission lines otherwise). The overdensity is computed as the number of detected galaxies over the number of expected galaxies in blank fields over the same volume (see \S~\ref{ssec:cross}). Column (7) show the overdensity of LAEs at $R\lesssim7\,h^{-1}\rm cMpc$ within $\pm1600\,\rm km\,s^{-1}$ in each field \citep{Garcia-Vergara2019}.}
\end{table*}

\begin{figure}
\centering
\includegraphics[height=7.8cm]{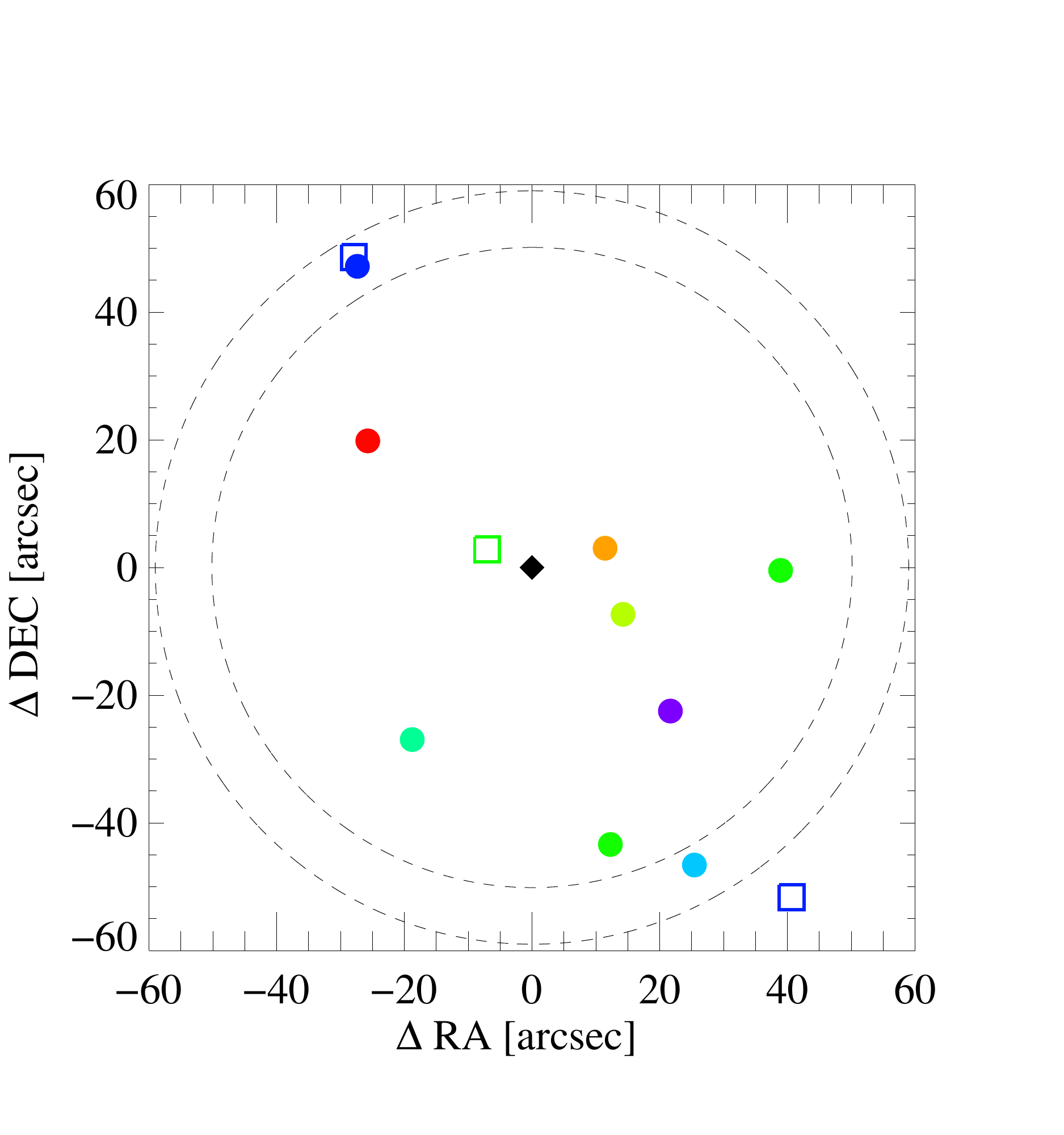} 
\caption{Sky distribution of the emission lines around the central quasar for our 17 fields combined. The central black diamond indicates the quasar position, the lines are indicated as filled circles with different colors indicating different fields in which they were detected. We include the position of the LAEs detected with $\rm S/N\geq5$ in the field as open squares. The biggest dashed line circle shows the region where we search for line candidates, which corresponds to the whole ALMA pointing (a radius of $59\arcsec$ from the central quasar) . The smaller dashed line circle shows the radius at which the telescope sensitivity is $\geq20\%$ of the maximum. Sources located outside of this limit radius were included in our catalog only if they have a fidelity $\geq$0.9. \label{fig:sky_dist}}
\end{figure}

\begin{figure}
\centering
\includegraphics[height=6.5cm]{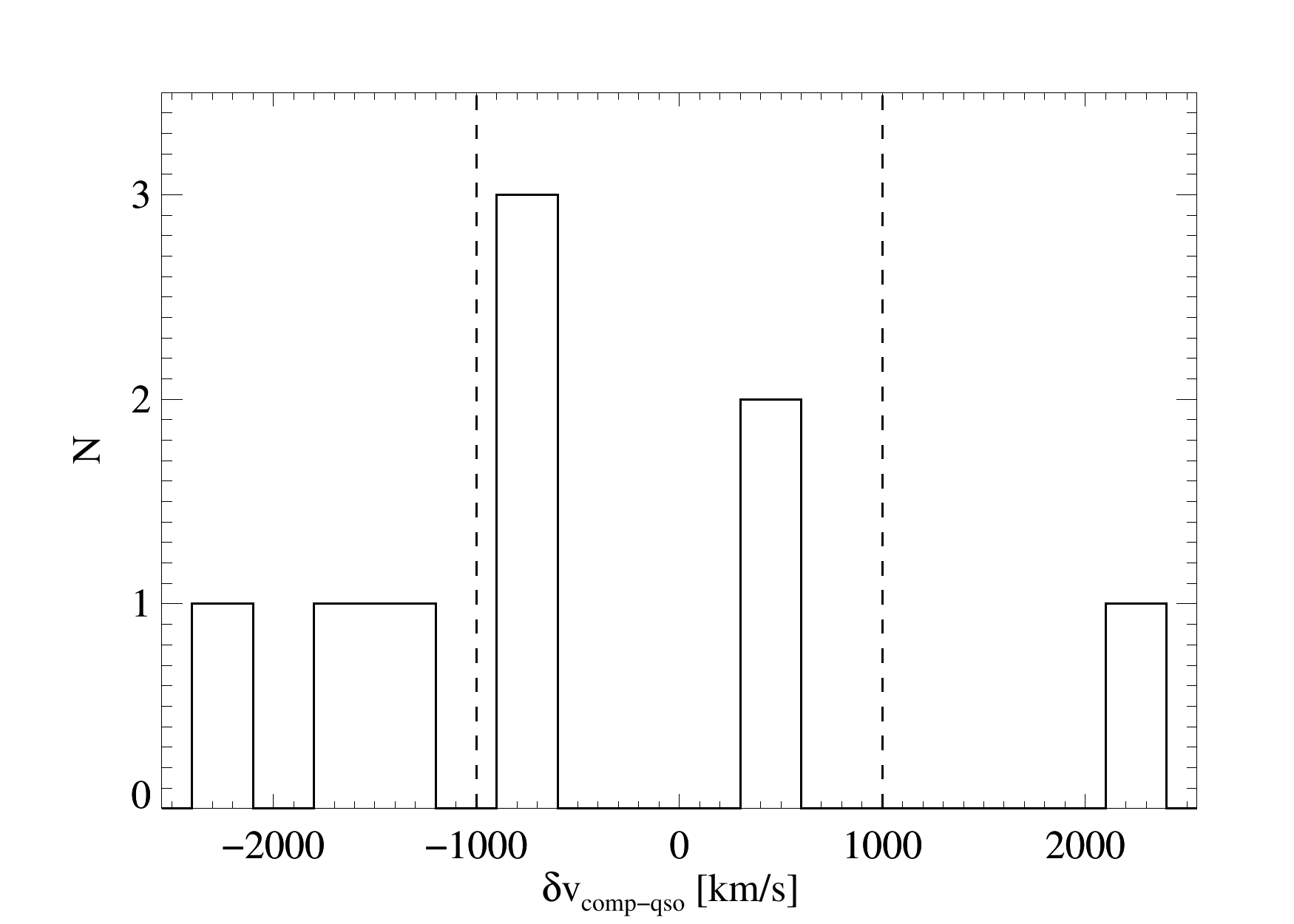} 
\caption{Velocity offset distribution of the CO(4--3) lines around the central quasar for our 17 fields together. The vertical dashed lines indicate the velocity window that was chosen to measure the clustering of sources around quasars (see \S~\ref{ssec:cross}). \label{fig:vel_dist}}
\end{figure}

In Fig.~\ref{fig:sky_dist} we show the sky distribution of the detected companion lines around the central quasar, and in Fig.~\ref{fig:vel_dist} we show the distribution of the velocity offsets between the lines and the central quasar for all our fields together. For the computation of the velocity offsets, we use the redshift based on the ALMA detection of the quasar, when it is detected with fidelity $\geq0.8$, otherwise, we use the redshift as determined from the optical quasar spectrum (reported in Table~\ref{table:quasar}).

\section{Clustering analysis}
\label{sec:clustering}

In this section, we measure the clustering properties of the CO(4--3) emitters around quasars, following an analogous procedure as the one in our previous studies about the clustering of LAE and LBG around $z\sim4$ quasars \citep{Garcia-Vergara2017,Garcia-Vergara2019}. We refer the reader to these works for detailed descriptions, but in \S~\ref{ssec:cross} we provide a brief overview. We also use the measured cross-correlation function to infer the clustering of CO emitters in blank fields in \S~\ref{ssec:auto}.  We note that in this study we do not analyze the clustering of continuum sources, since the clustering signal would be strongly diluted when projected over the large radial comoving distance traced by the almost flat selection function of these sources over the redshift range $1\lesssim z\lesssim7$.

\subsection{Quasar-CO cross-correlation function}
\label{ssec:cross}

We measure a volume-averaged projected cross-correlation function between quasars and CO(4--3) emitters defined by
\begin{equation}
\chi (R) = \frac{\int_{V_{\rm eff}} \xi(R,Z) dV}{V_{\rm eff}}
\label{eq:chi}
\end{equation}
where $\xi(R,Z)$ is the real-space correlation function, assumed to have a power-law shape $\xi(r) = (r/r_{\rm 0, QG})^{-\gamma}$, where $r_{\rm 0,QG}$ and $\gamma$ are the correlation length and the slope of the correlation function, respectively. The real-space separation between objects, $r$, has been written in eqn.~(\ref{eq:chi}) as a function of their two components: the transverse comoving separation $R$, and the radial comoving separation $Z$. $V_{\rm eff}$ is the effective volume of the survey, which is a cylindrical shell centered on each quasar, with inner and outer radius given by $R_{\rm min}$ and $R_{\rm max}$ respectively, and with a height defined by the radial comoving distance $Z_{\rm max}-Z_{\rm min}$.

We compute $\chi (R)$ in logarithmically spaced radial bins centered on the quasar by using the estimator
\begin{equation}
\chi (R) = \frac{\langle QG (R) \rangle}{\langle QR(R) \rangle} -1
\label{eq:chi_est}
\end{equation}
where $\langle QG (R) \rangle$ is the number of quasar-galaxy pairs observed in our survey, within the cylindrical shell volume, and $\langle QR (R) \rangle$ is the number of quasar-galaxy pairs that we expect to observe within the same volume but in the absence of clustering (i.e.\,assuming the background number density of CO emitting galaxies at $z\sim4$).

If we would have ALMA observations covering large areas of the sky at random locations (i.e.\,regions not containing quasars), using the same setup as for our data, we could directly measure $\langle QR (R) \rangle$ from the data. However, this is not the case. We cannot use the three quasar-free SPWs (SPWs 1,2, and 3)  for this purpose because i) the covered volume is small, resulting in extremely low statistics for the number counts\footnote{We computed the probability to detect a CO(4--3) line in each SPW, assuming the background number density from the CO(4--3) luminosity function measured in \citet{Decarli2019} (more details of this computation are provided in the main text). This probability is 5\%, 7\% and 7\% for SPW1, SPW2 and SPW3 respectively, which would result in a total of $\sim3$ CO(4--3) lines for the 17 fields detected in the volume covered by the three SPWs.}, ii) the central frequency (at least for SPW1) is relatively close ($<6000\,\rm km\,s^{-1}$) to the quasar location, and thus the background number counts could not necessarily be reached. The other two SPWs are far enough away, but they would be tracing the number density of CO(4--3) at lower redshift ($z\sim3.2$), and iii) the spectral resolution in SPW1, SPW2, and SPW3 is 2 times lower than the resolution in SPW0. We thus decided to estimate $\langle QR (R) \rangle$ using the CO(4--3) luminosity function measured at $z=3.8$ from the ASPECS survey \citep{Decarli2019}. We detail this in what follows. 

The number of quasar-galaxy pairs that we expect to observe in the absence of clustering is given by $\langle QR(R) \rangle = \int_{V_{\rm eff}} n_{CO}(L'\geq L'_{\rm Lim}) dV$, and for the cylindrical shell volume defined above, we can write
\begin{equation}
\langle QR(R) \rangle = \int_{Z{\rm min}}^{Z_{\rm max}} \int_{R{\rm min}}^{R_{\rm max}} n_{CO}(L'\geq L'_{\rm Lim}) 2\pi R dR dZ 
\label{eq:QR}
\end{equation}
where $n_{CO}(L'\geq L'_{\rm Lim})$ is the number density of CO(4--3) emitters with line luminosity above the limiting luminosity of our survey. To compute the  limiting luminosity, we follow \citet{Solomon1997},
\begin{equation}
\frac{L'_{\rm Lim}}{\rm K\, km\, s^{-1}pc^{2}} = \frac{3.257 \times 10^{7}}{1+z}\frac{F_{\rm Lim}}{\rm Jy\, km\, s^{-1}} \left(\frac{\nu_{0}}{\rm GHz}\right)^{-2} \left(\frac{D_{\rm L}}{\rm Mpc}\right)^{2}
\label{eqn:Lum}
\end{equation}
where $\nu_{0}$ is the rest-frame frequency of the line ($\nu_{0}=461.04$ for CO(4--3)), $D_{\rm L}$ is the luminosity distance at the redshift of the source, $z$, and $F_{\rm Lim}$ is the limiting integrated line flux. 

The limiting integrated line flux for each quasar field is computed using the rms noise of the cubes from Table~\ref{tab:obs}, and the $\rm S/N$ threshold defined by our fidelity threshold. Specifically, we assume a typical FWHM line width of $\rm 331\,km\,s^{-1}$, which is the median width measured in the high-fidelity CO line sample of the ASPECS survey\footnote{We note that this is larger than the median FWHM of the five sources used for our clustering analysis (median FWHM is $173\pm161$ for the five sources with fidelity $>$80\% and $301\pm183$ for the four sources with fidelity $>$90\%), but still consistent within error bars. We nevertheless checked that the assumed line width has only a little impact on our final clustering results.} \citep{Gonzalez-lopez2019}, and we use the corresponding $\rm S/N$ threshold for such a width ($\rm S/N=5.6$ for a detection with fidelity 0.8 as shown by the magenta curve in Fig.~\ref{fig:fidelity}). The limiting luminosity of the survey at the center of the pointing is reported in Table~\ref{tab:obs}. 

If the rms sensitivity of the survey were roughly flat, the term $n_{CO}(L'\geq L'_{\rm Lim})$ in eqn.~(\ref{eq:QR}) could be taken out from the integral and it can be simply computed by integrating the CO(4--3) luminosity function down to the limiting luminosity of the survey computed from eqn.~(\ref{eqn:Lum}). However, since the ALMA sensitivity decreases with increasing the radius within the pointing, the limiting luminosity actually depends on $R$, and therefore we have to model this dependency and numerically compute the integral over $R$ in eqn.~(\ref{eq:QR}). 

We model the sensitivity variation over the pointing as a 2D Gaussian with FWHM given by the primary beam size ($\sim66\arcsec$), and compute the primary-beam-corrected limiting integrated line flux as a function of $R$, which gives us the limiting luminosity as a function of $R$ from eqn.~(\ref{eqn:Lum}).  

Finally, the limiting luminosity as a function of $R$ is used to compute $n_{CO}(L'\geq L'_{\rm Lim})$ at every $dR$ integration step in eqn.~(\ref{eq:QR}). Here we use the CO(4--3) luminosity function at $z=3.8$ measured by \citet{Decarli2019}, which is given by the Schechter parameters $\rm log \Phi_{*} [Mpc^{-3} dex^{-1}]= -3.43$, $\rm log L'_{*} [K\, km\, s^{-1}pc^{2}]= 9.98$ for a fixed $\alpha=-0.2$. 

In principle, in eqn.~(\ref{eq:QR}) we should include a multiplicative term encapsulating the completeness of the sample. This completeness refers to the fraction of sources detectable by the search algorithm, and it is dependent on the line flux, line width, fidelity, and S/N of the sources\footnote{Note that this is not the completeness due to the sensitivity variation over the pointing, which is indeed corrected in our computation as explained above.}. Based on previous estimations of the performance of the line search algorithm \textsc{findclumps} using simulated mock sources \citep[e.g.][]{Gonzalez-lopez2019,Decarli2020,Loiacono2021}, we expect that at the high fidelity and S/N threshold used in our study, the completeness is near to 100\%, so we have neglected this completeness correction. This could result in a slight underestimation of the clustering of sources around quasars, but our main conclusions will not significantly change. 

\begin{figure}
\centering
\includegraphics[height=9.5cm]{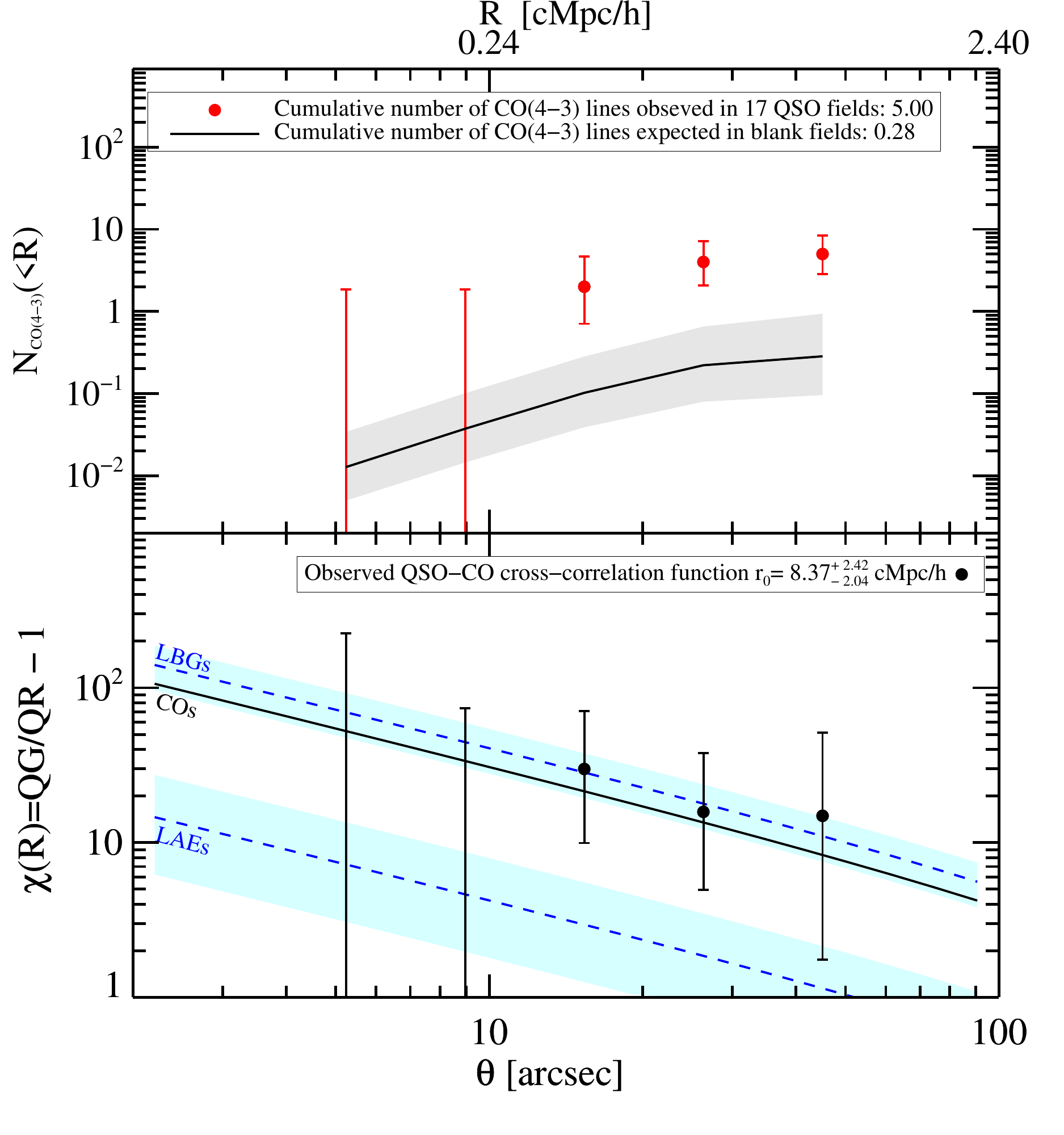}
\caption{\textit{Top:} The cumulative number counts of CO(4--3) lines observed in our 17 quasar fields  ($\langle QG(\leq R) \rangle$) within $\rm \Delta v = 1000\, \rm km\,s^{-1}$ from the central quasar (red points) with Poisson error bars, compared to the expectation for CO(4--3) lines in blank fields in our 17 quasar fields ($\langle QR(\leq R) \rangle$) computed using eqn.~(\ref{eq:QR}) (black line). The gray area shows the uncertainty in this expectation. Our observations yield 5 CO(4--3) lines from the ensemble of the 17 fields, while we expect only 0.28 CO(4--3) lines from the background alone, resulting in a total overdensity of $17.6^{+11.9}_{-7.6}$. \textit{Bottom:} Quasar-CO cross-correlation function $\chi (R)$ with $1\sigma$ Poisson error bars for the 17 fields (black data points) computed using eqn.~(\ref{eq:chi_est}), and the best maximum likelihood estimator for both $r_{\rm 0, QG}$, assuming a fixed $\gamma=1.8$ (black line). For comparison we show the quasar-LBG and quasar-LAE cross-correlation function (blue dashed lines) observed at $z\sim4$, and their respective uncertainties showed as blue shaded areas \citep{Garcia-Vergara2017,Garcia-Vergara2019}. \label{fig:CF}}
\end{figure}

The computation of $\langle QG(R) \rangle$ and $\langle QR(R) \rangle$ is performed individually for each of the 17 fields, and then they are stacked to obtain the final $\chi (R)$ value using eqn.~(\ref{eq:chi_est}). We measure the clustering using $Z_{\rm max} =  - Z_{\rm min} = 8.19\,h^{-1} \rm cMpc$  (corresponding to $\rm \Delta v = \pm1000\rm\, km\, s^{-1}$ at $z=3.87$). This choice results in a total of five sources included in the clustering analysis (see Table~\ref{table:line_prop}). Given the small size of our sample, we assume that Poisson error dominates our measurement, and we compute the one-sided Poisson confidence interval for small number statistics from \citet{Gehrels1986}. We show the measured cross-correlation in Fig.~\ref{fig:CF} and tabulate it in Table~\ref{table:CF}. 

\begin{table}
\caption{Quasar-CO cross correlation function. \label{table:CF}}
\centering
\begin{tabular}{l l r r l} 
\hline
$R_{\rm min}$ & $R_{\rm max}$ & $\langle QG(R)\rangle$ & $\langle QR(R)\rangle$ & $\chi (R)$\\
$h^{-1} \rm cMpc$ & $h^{-1} \rm cMpc$ & &   &  \\
\hline 
  0.096 &   0.165 &   0.000 &   0.008 &  -1.00$^{+226.13}_{-  0.00}$ \\
  0.165 &   0.282 &   0.000 &   0.025 &  -1.00$^{+ 74.91}_{-  0.00}$ \\
  0.282 &   0.483 &   2.000 &   0.065 &  29.89$^{+ 40.75}_{- 19.96}$ \\
  0.483 &   0.827 &   2.000 &   0.119 &  15.79$^{+ 22.15}_{- 10.85}$ \\
  0.827 &   1.417 &   1.000 &   0.063 &  14.88$^{+ 36.53}_{- 13.14}$ \\
\hline
\end{tabular}
\end{table}

Note that choosing a larger radial comoving range ($\rm \Delta v$) would result in more sources for the measurement, increasing the statistics, but the clustering signal would be diluted because we are integrating the signal in a larger volume and up to large distances from the quasar, where the background is close to being reached. We also note that keeping the volume small also decreases the probability of having a contaminant in the sample (see \S~\ref{table:contam}). We explore the effect of the chosen $\rm \Delta v$ in Appendix \S~\ref{sec:impact}.

For the seven quasars not (securely) detected in our observations, we use the optical-based redshifts, which could have an offset from the ALMA-based redshift of $\sim 700-800\rm\, km s^{-1}$ (see Fig.~\ref{fig:vel_offset_qso}). This would mean that for these quasars we may be including or excluding companions at larger and smaller redshift distances, respectively. The confirmation of the precise redshift for these quasars is the only way to correct for these uncertainties in our clustering measurement, but we explore how much the clustering signal changes if we use ALMA-based redshifts for all the quasars, even if they are detected at low fidelity. We find that the same five CO emitters are kept in this case, confirming the stability of the obtained correlation function.

To determine the real-space cross-correlation parameter $r_{\rm 0,QG}$ that best fits our data, we use a Poisson maximum likelihood estimator. Given the noise of the data and the small physical scales proved in our study, we follow common practice and fit our data assuming a fixed slope $\gamma=1.8$ \citep[e.g.][]{Ouchi2004,Garcia-Vergara2017,Garcia-Vergara2019,Fossati2021}. We find that the maximum likelihood and $1\sigma$ confidence interval for the correlation length is $r_{\rm 0,QG} =  8.37^{+2.42}_{-2.04}\,h^{-1} \rm cMpc$. We use the best fitted parameter in eqn.~(\ref{eq:chi}) to compute the corresponding $\chi(R)$ value, which is shown as a black line in Fig.~\ref{fig:CF}.

Finally, we use our $\langle QG(R) \rangle$ and $\langle QR(R) \rangle$ binned values to compute the observed and expected cumulative number counts of CO(4--3) lines in our 17 quasar fields ($\langle QG(\leq R) \rangle$), and show this in Fig.~\ref{fig:CF}. In the whole volume survey ($1751.7 \,h^{-3} \rm cMpc^{3}$ for the 17 fields over $\rm \Delta v = 1000\, \rm km\,s^{-1}$), we find a total of 5 CO(4--3) lines while we expect only 0.28 CO(4--3) lines from the background alone, resulting in a total CO(4--3) line overdensity of $17.6^{+11.9}_{-7.6}$ in quasar fields. We also compute the total overdensity per field as the ratio of $\langle QG(R) \rangle$ per field over $\langle QR(R) \rangle$ integrated over the radial bins and provide these values in Table~\ref{table:n_per_field}. Although the individual overdensity is dominated by low number statistics and affected by cosmic variance, we include this to study possible correlations between the overdensities and the quasar properties (Garc\'ia-Vergara et al. in preparation).

\subsection{Comparison with the Clustering of Other Populations Around Quasars}
\label{ssec:compar}
We compare our results with the clustering of LBG and LAE around $z\sim4$ quasars. Since the previous LBG and LAE clustering studies extend up to larger scales ($R\lesssim9\,h^{-1}\rm cMpc$) than the traced in our study ($R\lesssim1.5\,h^{-1}\rm cMpc$), for the comparison we assume that the small-scale quasar-CO cross-correlation function can be extrapolated towards larger scales following a single power-law shape. 

We find that the QSO-CO cross-correlation length is slightly lower, but consistent within error bars with the QSO-LBG cross-correlation length, which is given by $r_{\rm 0,Q-LBG} =  9.78^{+1.68}_{-1.86}\,h^{-1} \rm cMpc$ for a fixed $\gamma=1.8$\footnote{\citep{Garcia-Vergara2017} fitted the quasar-LBG cross-correlation function using a fixed $\gamma=2.0$, thus we re-fitted their measurements using a fixed $\gamma=1.8$, which results in $r_{\rm 0,Q-LBG} = 9.78^{+1.68}_{-1.86}\,h^{-1} \rm cMpc$.} \citep{Garcia-Vergara2017}. This suggests that CO emitters and LBGs would inhabit dark matter halos of similar masses at $z\sim4$ (we further discuss this point in \S~\ref{ssec:auto}). 

We can also compare our measurements with the clustering of LAEs around this same quasar sample, thus providing a direct comparison of optical and dusty galaxy populations in these fields. We find that the cross-correlation length for CO emitting galaxies is 3 times higher than the cross-correlation length for LAEs ($r_{\rm 0,Q-LAE} = 2.78^{+1.16}_{-1.05}\,h^{-1} \rm cMpc$ with $\gamma=1.8$; \citealt{Garcia-Vergara2019}) around quasars. We note that the redshift window traced by both studies is similar ($\rm \Delta v=\pm1000\rm\, km\, s^{-1}$ for CO emitters, and $\rm \Delta v=\pm1600\rm\, km\, s^{-1}$ for LAEs) and thus we do not expect that this discrepancy is the result of a dilution in the signal due to differences in the traced volume. We have explicitly checked this in Appendix \S~\ref{sec:impact}, where we find that the quasar-CO cross-correlation length is only 1.1 times smaller if we assume a $\rm \Delta v=\pm1600\rm\, km\, s^{-1}$ (see Fig.~\ref{fig:impact_vel}).

Tracing a different quasar sample at $z=3-4.5$, and using deeper optical observations, \citep{Fossati2021} measure a slightly higher quasar-LAE cross-correlation length of $r_0=3.15^{+0.36}_{-0.40}\,h^{-1}\rm cMpc$ at a fixed slope $\gamma=1.8$\footnote{The $r_0$ value published in \citet{Fossati2021} has been recently re-computed. Here we quote the most updated $r_0$ value, which was kindly provided by the authors by private communication.}. This is still 2.7 times lower than the cross-correlation length for CO emitting galaxies. We note that the physical scale traced in this study ($R\lesssim0.6\,h^{-1}\rm cMpc$) is slightly smaller to that traced in our analysis.

The difference in the clustering of CO emitters and LAEs around quasars is unlikely to be caused by differences in the halo mass hosting both populations (see \S~\ref{ssec:auto}), thus we suggest that this discrepancy is related to physical processes affecting the visibility of the LAEs around quasars. We further discuss this interpretation in section \S~\ref{sec:discussion}.

\subsection{Auto-correlation of CO emitters at $z\sim4$}
\label{ssec:auto}

The auto-correlation of a galaxy population is a powerful tool, since it can be directly related to the dark matter halo mass in which that population reside \citep[e.g.][]{Cole1989,Mo1996,Sheth1999}. The auto-correlation of CO emitting galaxies at $z\sim4$ has never been measured before, mainly due to the lack of large and deep surveys of these galaxies at high$-z$. The largest samples of CO emitters at $z\sim4$ currently available are composed of only a few tens of sources \citep[e.g.][]{Decarli2016,Decarli2019}, which do not provide enough statistics for an auto-correlation measurement. Therefore, to-date it has been challenging to compute precise masses of CO emitting galaxies, and understand how they fit into different evolutionary scenarios for high$-z$ galaxies.

However, under certain assumptions, the cross-correlation between CO emitters and quasars can be used to infer the clustering of CO emitting galaxies in blank fields, providing first constraints on the clustering of this population at $z\sim4$. 

First, we assume that our small-scale cross-correlation can be extrapolated towards larger scales following a power-law shape given by $\xi(r) = (r/r_{\rm 0,QG})^{-\gamma}$. Although the auto-correlation function of quasars and galaxies has been found to slightly deviate from a power-law towards smaller ($\lesssim0.2\,h^{-1}\rm cMpc$) scales \citep[e.g.][but see also \citealt{Shen2010}]{Hennawi2006,Ouchi2005}, likely due to the transition between the one-halo to two-halo terms, the auto-correlation is typically reasonably well approximated as a power-law, and thus we assume that the one-halo term does not strongly boost the signal in our measurement. 

Second, we assume a deterministic bias model, in which the QSO-galaxy cross-correlation function can be written as $\xi_{\rm QG} = \sqrt{\xi_{\rm QQ}\xi_{\rm GG}}$, where $\xi_{\rm QQ}$ and $\xi_{\rm GG}$ are the auto-correlation of quasar and galaxies, respectively. We also assume that $\xi_{\rm QQ}$ and $\xi_{\rm GG}$ have a power-law shape with the same slope $\gamma=1.8$. Under these assumptions, the correlation lengths can be related by 
\begin{equation}
r_{\rm 0,QG} = \sqrt{r_{\rm 0,QQ}r_{\rm 0,GG}}
\label{eq:r0cross}
\end{equation}

Using the quasar auto-correlation length previously reported at $z\sim4$, given by $r_{\rm 0,QQ}= 22.3\pm 2.5\,h^{-1} \rm cMpc$ (re-computed from \citealt{Shen2007} with a fixed $\gamma=1.8$), and our measured QSO-galaxy cross-correlation length, we obtain that the auto-correlation length of CO emitters at $z\sim4$ is given by $r_{\rm 0,GG}= 3.14\pm1.71\,h^{-1} \rm cMpc$. 

This measurement is slightly lower than the LBG auto-correlation length at $z\sim4$ ($r_{\rm 0, LBG} = 4.1^{+0.2}_{-0.2}\,h^{-1} \rm cMpc$; \citealt{Ouchi2004}), and is slightly higher than the LAE auto-correlation length at $z\sim4$ ($r_{\rm 0, LAE} =  2.74^{+0.58}_{-0.72}\,h^{-1} \rm cMpc$, \citealt{Ouchi2010}). We caution that the uncertainties of our estimation are still large, which makes the CO auto-correlation length consistent within uncertainties with the auto-correlation length of both LAE and LBG. 

Our results suggest that CO emitters would inhabit dark matter halos with similar masses as these hosting LBGs and LAEs. Specifically, we calculate the bias factor (the clustering of CO emitters relative to the clustering of the underlying dark matter) using
\begin{equation}
b(z) = \left(\frac{r_{\rm 0,GG}}{8[h^{-1} \rm cMpc]}\right)^{(\gamma/2)}\frac{J_{2}^{1/2}}{\sigma_8D(z)/D(0)}
\label{eq:bias}
\end{equation}
\citep[][]{Peebles1980}, where $D(z)$ is the growth factor at redshift $z$ \citep{Carroll1992}, and $J_2=72/[(3-\gamma)(4-\gamma)(6-\gamma)2^\gamma]$. At $z=3.87$, this yields a bias value  $b=2.8^{+1.34}_{-1.42}$. Using the \citep{Sheth2001} formalism, our result implies a halo mass for CO emitters of $M_{\rm halo}= 8.31^{+49.29}_{-8.27} \times 10^{10} M_{\odot}$.

Interestingly, the reported halo masses for CO emitters, are lower than the median halo mass hosting $S_{870\rm \mu m}>1.2\rm\,mJy$ submillimeter galaxies (SMGs) detected with ALMA at $1<z<3$ ($M_{\rm halo} > 3.2 \times 10^{11} M_{\odot}$; \citealt{Garcia-Vergara2020}). If the SMG clustering does not strongly evolve with redshift as suggested by \citealt{Stach2021} (but see \citealt{Wilkinson2017}), then this would mean that CO emitters are lower mass galaxies compared to the population of SMGs typically detected in continuum surveys.  

While our measurements are still noisy, they provide a first rough constraint of the clustering of CO emitting galaxies, allowing us for the first time to locate them within the context of evolutionary galaxy models. Larger and deeper surveys of emitting lines around quasars are still needed to constrain the clustering of CO emitters more precisely. Alternatively, large and deep surveys of CO emitting galaxies in blank fields would offer an independent and more direct constraint of the CO clustering at these redshifts. However, such large surveys are very expensive and extremely challenging because of the required sensitivity and the low number density of CO emitters in blank fields.

\section{Discussion}
\label{sec:discussion}

Our study reveals a large overdensity of CO emitting galaxies ($17.6^{+11.9}_{-7.6}$) and a strong clustering of them around quasars ($r_{\rm 0,QG} = \rm 8.37^{+2.42}_{-2.04}\,h^{-1} \rm cMpc$) at scales $R\lesssim1.5\,h^{-1}\rm cMpc$. This result helps clarify the current confused picture of quasar environments at high$-z$ and provides strong observational evidence in favor of $z\sim4$ quasars as tracers of massive structures. By comparing with the previous measurement of clustering of LAEs around the same quasar sample at scales $R\lesssim7\,h^{-1}\rm cMpc$, we find that CO emitting galaxies are significantly more clustered than LAEs around quasars (with a cross-correlation length 3 times higher), resulting in large overdensities of these galaxies, whereas only a mild overdensity of LAEs ($1.4^{+0.4}_{-0.4}$) was found in these fields \citep{Garcia-Vergara2019}. The comparison with the other available LAE study in quasar environments \citep{Fossati2021} which is performed at smaller physical scales ($R\lesssim0.6\,h^{-1}\rm cMpc$), aims to similar conclusions, with the  quasar-CO cross-correlation length being 2.7 times higher than the quasar-LAE cross-correlation length (see \S~\ref{ssec:compar}). In the following, we discuss the possible reasons that could explain this discrepancy.

First, we explore the possibility that CO emitting galaxies inhabit more massive dark matter halos compared to LAEs, which would result in a significant difference in the quasar-galaxy cross-correlation for both populations. Although the dark matter halos for CO emitters at $z\sim4$ have not been constrained yet, the LAE auto-correlation length at $z\sim4$ is well constrained and is found to be $r_{\rm 0, LAE} = 2.74^{+0.58}_{-0.72}\,h^{-1} \rm cMpc$, \citep{Ouchi2010}. If we focus on clustering hierarchy arguments only and assume a deterministic bias model, the difference of a factor of 3 in the measured quasar-CO and quasar-LAE cross-correlation length, implies that the auto-correlation length of  CO emitters would be 9 times larger than the auto-correlation length of the LAEs (see eqn.~\ref{eq:r0cross}), resulting in $r_{\rm 0, CO} \sim 25 \,h^{-1} \rm cMpc$. 

This is even higher than the quasar auto-correlation length at $z\sim4$ ($r_{\rm 0,QQ}= \rm 22.3\pm 2.5\,h^{-1} \rm cMpc$, \citealt{Shen2007}, which implies halo masses of $M_{\rm halo} > 6 \times 10^{12} M_{\odot}\,h^{-1}$), and it would imply that CO emitters inhabit halos more massive than quasars. This seems to be an unlikely scenario and inconsistent with the CO dark matter halo mass inferred from our results, assuming a deterministic bias model ($M_{\rm halo}= 8.31^{+49.29}_{-8.27} \times 10^{10} M_{\odot}$, see \S~\ref{ssec:auto}).

Another possibility to explain the discrepancy in the overdensity of CO emitters and LAEs around quasars is related to the uncertainties on the rest-frame UV quasar redshift of our targets. If these quasar redshifts were associated with large uncertainties, the LAE search would have been offset from the real quasar position, which would result in a lower number density of detected galaxies, and thus a lower quasar-LAE cross-correlation. However, with our ALMA observations, we could confirm the redshift of 10 quasars (and tentatively for the other 7 quasars detected with lower fidelity), and for them we find relatively low redshift offset compared with the ones determined from rest-frame UV emission lines ($|\rm \delta v|=738\rm\,km\,s^{-1}$, see \S~\ref{ssec:final_catalogs}). These uncertainties in the quasar redshifts are much smaller than the FWHM of the narrow band used in the LAEs study ($3197\, \rm km\,s^{-1}$), implying that the LAE search was performed at the correct redshift. 

If we assume that the remaining seven quasars have similar redshift uncertainties (as indeed suggested by their lower fidelity detections, see Table~\ref{table:n_per_field}), then this is not a convincing explanation for the observed discrepancy. Additionally, we note that five out these seven quasars exhibit an overdensity of LAEs (see Table~\ref{table:n_per_field}) which is an indication that the LAE search was performed at the correct redshift in these cases. We note that one of the quasars exhibits a larger offset ($|\rm \delta v|=2386\rm\,km\,s^{-1}$ for J1224+0746), in which case the LAEs search was performed at larger distances from the quasar, where the background number density would be expected. However, if we exclude that field in the LAE study, we find that the total overdensity increases from $1.36$ to $1.43$ and the auto-correlation length increase by a factor of 1.09, which would not change our main conclusions. 

Ruling out the two mentioned explanations as the main reasons to explain the large overdensity of CO emitters compared to the mild overdensity of LAEs in the same fields, we explore the possibility that particular physical properties in galaxies around quasars could be impacting their visibility and detection. While the CO(4--3) emission line traces the molecular gas in a galaxy\footnote{The luminosity of the CO(4--3) emission line $L'_{\rm CO(4-3)}$ can be converted to a molecular hydrogen gas mass $M_{\rm mol}$ by assuming an excitation correction $L'_{\rm CO(4-3)}/L'_{\rm CO(1-0)}$ and a conversion factor $\alpha_{\rm CO}$ \citep[e.g.][]{Carilli2013}.}, the Ly$\alpha$ emission line is a tracer of instantaneous star formation, and thus, a relatively small star formation efficiency in galaxies around quasars could explain the lack of LAEs in these fields. 

Although the molecular gas reservoirs in galaxies have been found to correlate well with their star formation rate, and stellar mass \citep[e.g.][]{Kennicutt1998,Bigiel2008,Leroy2008}, these scaling relations have been mostly measured in field galaxies, but the validation of such relations in high-$z$ dense environments is still poorly constrained. The few available studies have been aimed to trace the properties of galaxies in $z\sim1.5$ galaxy clusters, and they show evidence of systematic deviations from the scaling relations of the field towards larger molecular gas masses \citep[e.g.][but see also \citealt{Rudnick2017}]{Noble2017,Hayashi2018}. Although such systems have a much dense and evolved intercluster medium than that in protoclusters at $z\sim4$, the results of these studies suggest that complex physical processes may be involved in the galaxy evolution in dense environments. 

We note, however, that the Ly$\alpha$ emission does not depend only on the instantaneous star formation, but it also depends on other factors such as the inter-stellar and circum-galactic medium conditions, which impact the complex radiative transfer process of the Ly$\alpha$ photons. Exploring the mentioned scenario in the context of our galaxy samples would require the acquisition of more observations. Specifically, multi-wavelength photometry would be useful to constrain the properties of the galaxies, including SFRs. Additionally, James Webb Space Telescope (JWST; \citealt{Gardner2006}) observations can add key information of kinematics, stellar masses, and SFRs which would help to build a physical scenario for these systems. Finally, ALMA [CII] 158-$\mu m$ emission observations are sensitive tracers of the interstellar medium conditions, enabling the study of scaling relations in high$-z$ dense environments. 

Finally, as previously suggested by \citet{Garcia-Vergara2019}, galaxies around quasars could be more dusty, affecting the visibility of the Ly$\alpha$ line. Since Ly$\alpha$ photons are easily absorbed by dust, the Ly$\alpha$ emission line could be partially suppressed, becoming hardly detectable. This would result in a decreased number of detected LAEs around quasars. The CO emission line is instead unaffected by dust absorption, making it detectable even in galaxies with large fractions of dust. This could explain why the deeper LAE study \citep{Fossati2021} reveals a stronger clustering of LAEs around high$-z$ quasars compared to that computed from shallower LAEs observations \citep{Garcia-Vergara2019}.

We note that in this scenario the detection of LBGs would also be affected by the dust, implying a lower LBG overdensity around quasars than that expected. However, the escape of the Ly$\alpha$ photons is particularly sensitive to the presence of dust, resulting in a strongly attenuated Ly$\alpha$ emission line, whereas the UV-continuum may be less impacted, making LBGs still detectable. ALMA continuum observations of our quasar fields would allow us to directly trace the dust in these galaxies, and together with additional multi-wavelength data, it would provide invaluable information to completely characterize the galaxy properties through SED modeling, testing their dependence on environment.

We caution that our clustering constraints are still dominated by the low number statistics provided by our relatively small quasar sample. Deeper and/or larger surveys of quasar fields would constrain the galaxy clustering around quasars with higher signal-to-noise than that achieved in this work. Alternatively, targeting a brighter emission line, such as the emission line [CII] at 158-$\mu m$ (which is the strongest line from star forming galaxies at radio wavelengths; \citealt{Carilli2013}), would offer better statistics with the same exposure time despite of the smaller area ($R\lesssim0.3\,h^{-1}\rm cMpc$) that we would trace using ALMA/band 8 at which this line is detectable at $z=3.87$. Specifically, we computed that for the same exposure time used in our CO emitters study, we expect to detect 17 [CII] companions for all the quasar fields over the area covered with band 8\footnote{For this computation we have used the [CII] luminosity function from \citet{Loiacono2021}, and assumed that the quasar-CII cross-correlation function is the same as the quasar-CO cross-correlation function at $z\sim4$.}. Finally, we stress the importance of performing quasar environment studies using multi-wavelength information, since the visibility of the galaxies around quasars seems to be strongly dependent on the studied wavelengths. 

\section{Summary}
\label{sec:summary}
We use ALMA band 3 observations to perform a blind search for CO(4--3) emitting galaxies in the environment of 17 $z\sim4$ quasars. The quasars were selected from the SDSS and BOSS quasar catalogs to lie within a redshift window of $z\sim3.862-3.879$ set by an optical narrow-band filter used in a previous work to detect LAEs around the quasars. The spectroscopic redshifts of the quasars are determined from the UV emission lines (with typical uncertainties of $<800\rm\,km\,s^{-1}$). 

We explore a cylindrical volume around the quasars defined by a projected radius of $R\lesssim1.5\,h^{-1}\rm cMpc$ and a velocity coverage of $\rm \Delta v \sim \pm 3000\, \rm km\,s^{-1}$, and find 9 CO-emitting galaxies with $\rm S/N\geq5.6$, and line widths ranging from $\rm 47\,km\,s^{-1}$ to $\rm 728\,km\,s^{-1}$. The S/N threshold was chosen to only select sources with fidelity $\geq0.8$ in our survey. 

We also detect the CO(4--3) emission line from 10 quasars with fidelity $\geq0.8$, and they typically show relatively low redshift offsets (the median offset is $|\rm \delta v|=738\rm\,km\,s^{-1}$) with respect to the redshifts determined from the UV-emission lines, except for one quasar which exhibit an offset of $|\rm \delta v|=2386\rm\,km\,s^{-1}$. The other 7 quasars were tentatively detected with lower fidelity ($<0.4$), and S/N ranging from 4.1 to 5.3.

We quantify the clustering properties of the galaxies around the quasars. For this, we only focus on a small velocity range of $\rm \Delta v \sim \pm 1000\, \rm km\,s^{-1}$ around the quasar, to avoid dilution of the small-scale clustering signal. Five of the CO emitting galaxies lie within this volume, and we use this sample to measure the volume-averaged cross-correlation function. For this measurement, we estimate the background number density from the CO luminosity function previously measured at $z=3.8$ from blank fields \citep{Decarli2019}. 

We find that the expected number density of CO(4--3) emitting galaxies in blank fields is 0.28 for the whole sample of quasars, while 5 were detected in our survey, which results in an overdensity of $17.6^{+11.9}_{-7.6}$. We also fit the quasar-CO cross-correlation in our fields and find a cross-correlation length of  $r_{\rm 0,QG} =  8.37^{+2.42}_{-2.04}\,h^{-1} \rm cMpc$, assuming a fixed slope $\gamma=1.8$.

In a previous study we performed a search for LAEs in the same 17 quasar fields, which allow us to simultaneously trace the clustering properties of optical and dusty galaxies around quasars for the first time. Such a study revealed only a mild overdensity (x1.4) of LAEs in these fields, and a quasar-galaxy cross-correlation length of $r_{\rm 0,Q-LAE} =  2.78^{+1.16}_{-1.05}\,h^{-1} \rm cMpc$ \citep{Garcia-Vergara2019}, which is 3 times lower than the cross-correlation length found for CO emitters. 

We argue that differences in the halo mass hosting the two galaxy populations, and uncertainties associated with the optically-based quasar redshifts are unlikely reasons to explain the observed discrepancy. We suggest instead that the properties of galaxies in quasar environments could impact their visibility and thus detectability in Ly$\alpha$ emission. Specifically, galaxies in quasar environments could have low star formation efficiency or could have an excess of dust, becoming more difficult to detect in Ly$\alpha$ emission. Exploring these mentioned scenarios would be only possible with the acquisition of additional multi-wavelength observations.  

Finally, we use our quasar-CO cross-correlation to infer the clustering of CO emitting galaxies at $z\sim4$. Assuming a deterministic bias model, and extrapolating the observed small-scale cross-correlation up to larger scales, we find an auto-correlation length for CO emitters of $r_{\rm 0,CO} =  3.14\pm1.71\,h^{-1} \rm cMpc$ (for a fixed slope $\gamma=1.8$), which agrees well, within the $1\sigma$ error bars, with the clustering of LBG and LAE at $z\sim4$, but is lower than the clustering of SMGs at $1<z<3$. Assuming that there is not a strong evolution of the clustering of SMGs with redshift, this would imply that the population of CO emitters inhabit less massive halos compared to the general population of  SMGs. Larger surveys of CO emitters are required to have an independently constrain their auto-correlation.

As the first quasar sample targeted for clustering studies of both optical and dusty galaxies, our study demonstrates the importance of tracing different galaxy populations, and it also opens new questions about environmental effects on galaxy evolution, highlighting the importance of characterizing galaxies in the vicinity of high$-z$ quasars. 

\appendix
\section{Impact of the Fidelity and $\rm \Delta v$ Choice on the Cross-Correlation Function}
\label{sec:impact}
In this Appendix, we explore how the Fidelity and $\rm \Delta v$ choices impact the cross-correlation measurement that we present in section \S~\ref{sec:clustering}. 

\begin{figure}
\includegraphics[height=6.4cm]{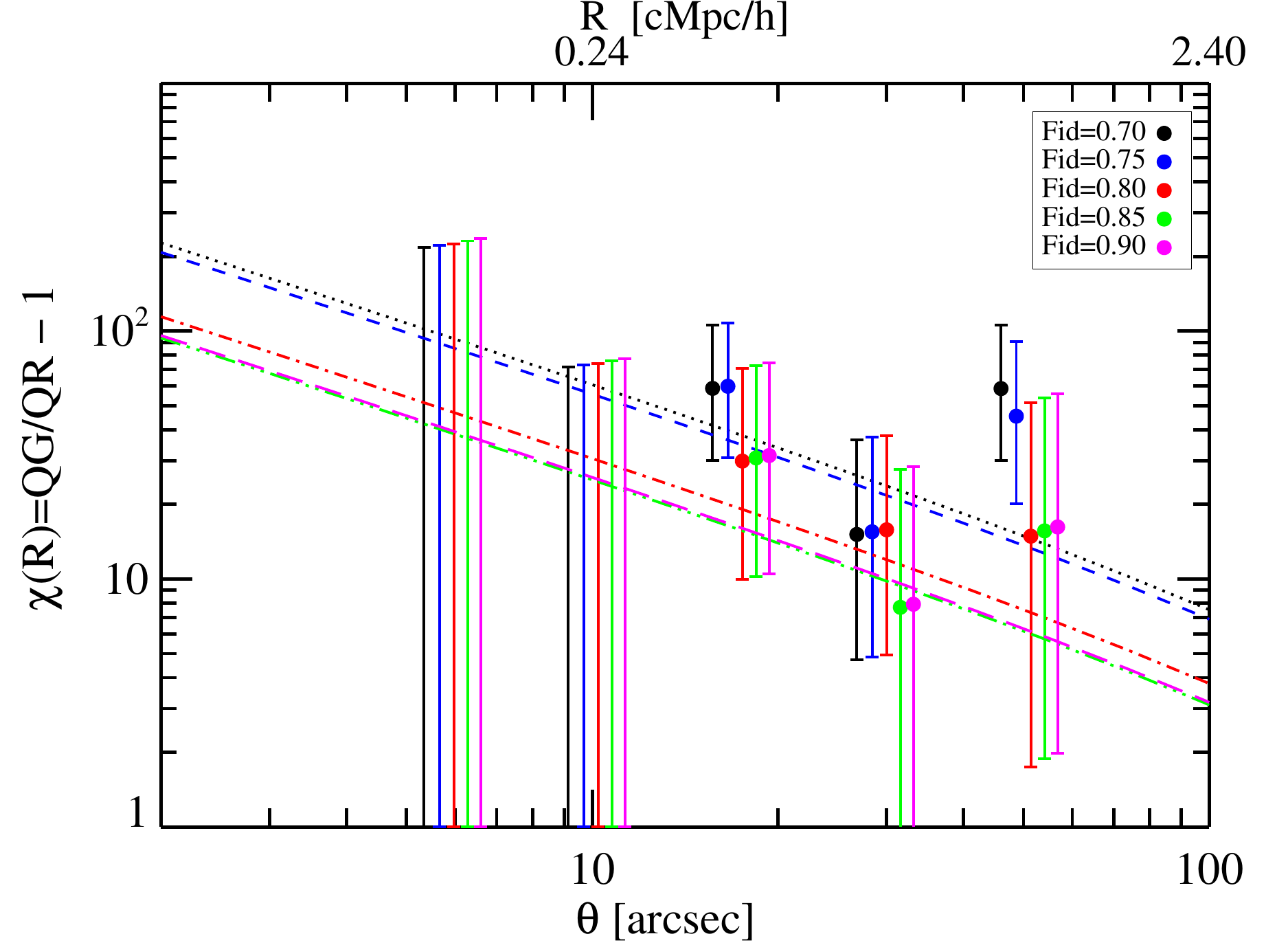}
\caption{Impact of the fidelity threshold adopted to create the catalog of sources. The results converge at fidelity$\geq0.8$. The clustering signal becomes flat when including sources with lower fidelities, which means that we are including contaminants (i.e.\,noise fluctuations being detected as real sources by the algorithm). \label{fig:impact_fid}}
\end{figure}

We first check the impact of the fidelity choice on the measurement. For this, we create source catalogs with different fidelity thresholds ranging from 0.7 to 0.9 in steps of 0.05 (to follow the same criteria as that described in \S~\ref{ssec:final_catalogs}, in all the cases we increase the fidelity threshold by 0.1 for sources selected at distances larger than 50.14\arcsec\ from the central quasar, corresponding to the radius at which the telescope sensitivity is $\geq20\%$ of the maximum). We compute the cross-correlation function following the same procedure described in \S~\ref{ssec:cross}, and fit the correlation with a fixed slope $\gamma=1.8$. We show our results in Fig.~\ref{fig:impact_fid}. We find that the cross-correlation decreases with increasing fidelity threshold, and it converges for fidelity$\geq0.8$, which motivates the choice of this as the threshold for our study. 

We also note that at lower fidelity ($<0.8)$ we include more sources, but most of them are impacting the last bin of the measurement (at $\theta\sim50\arcsec$), making the cross-correlation flatter, with the fixed slope at $\gamma=1.8$ poorly constraining their shape. This suggests that we are including contamination in these samples (i.e.\,probably fake noise fluctuations that are detected as real sources by the line search algorithm), which dilute the power-law signal, causing the flattening. 

\begin{figure}
\includegraphics[height=6.4cm]{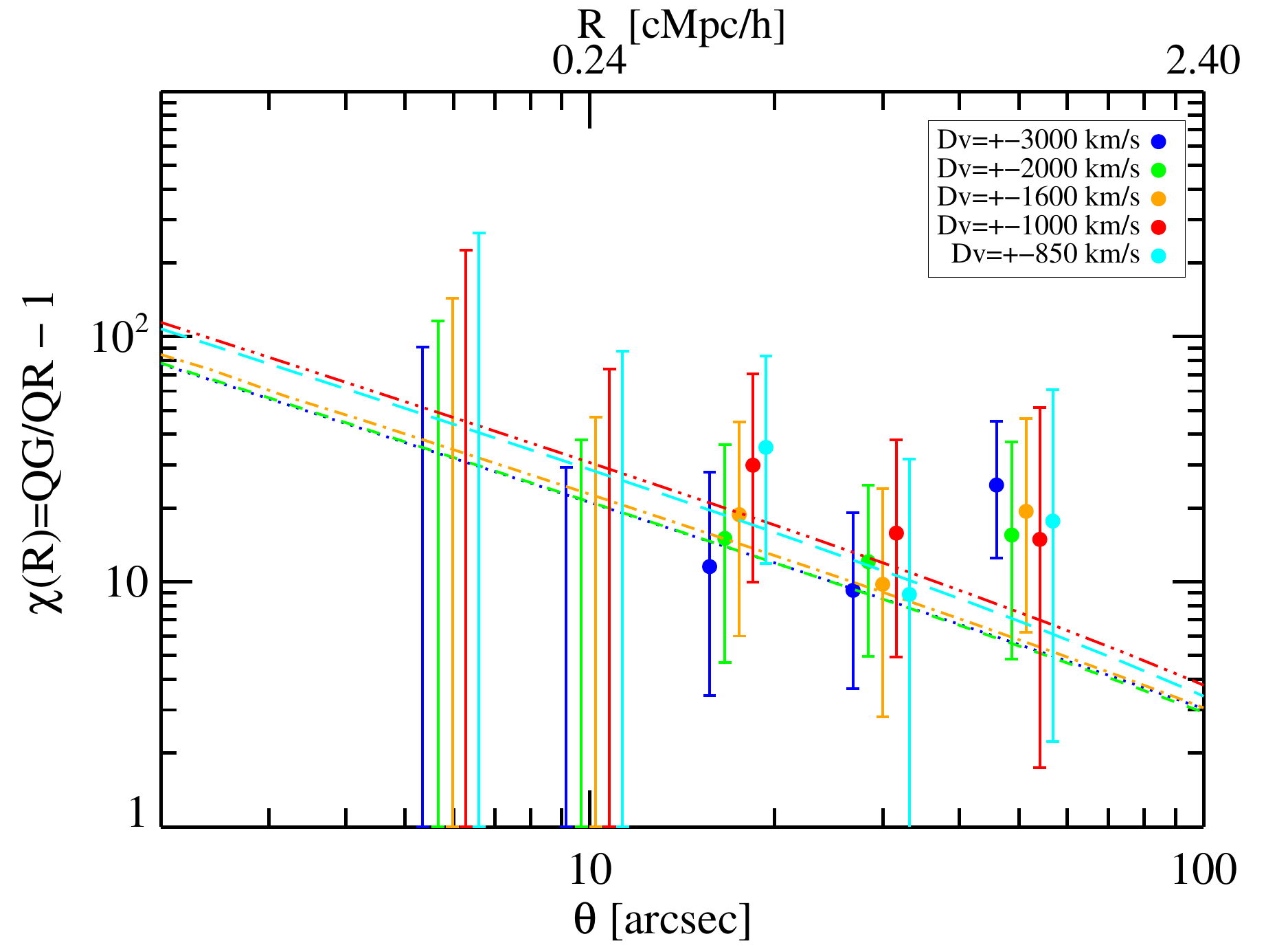}
\caption{Impact of the radial comoving range around the quasar ($\rm \Delta v$) used to measure the clustering. The clustering slightly decreases when increasing the $\rm \Delta v$, because small-scale signal starts to be diluted when integrating over a large volume, covering distances from the quasar at which the background number density is close to being reached. \label{fig:impact_vel}}
\end{figure}

We also explore the impact of the radial comoving range ($\rm \Delta v$) choice. For this, we only focus on the sources with fidelity $\geq0.8$ (i.e.\,on the 9 sources of Table~\ref{table:line_prop}), and compute the correlation-function using different $\rm \Delta v$ values ranging from $850\,\rm km\, s^{-1}$ to $3000\,\rm km\, s^{-1}$ (i.e.\,the whole cube). We show our results in Fig.~\ref{fig:impact_vel}. As mentioned in \S~\ref{ssec:cross}, the choice of a larger radial comoving range increases the statistics, reducing the error bars in the measurement, but it also dilutes the strong small-scale signal, because we are integrating over larger volumes and at large distances from the quasar where the number counts of the background start to be reached. This effect is relatively small, but can be seen to happen at $\rm \Delta v\geq1600\,\rm km\, s^{-1}$. Our choice of $\rm \Delta v=1000\,\rm km\, s^{-1}$ (corresponding to $8.19\,h^{-1} \rm cMpc$ at $z=3.87$) is a balance between having good statistics, but not strongly diluting the signal.

\begin{acknowledgements}
This paper makes use of the following ALMA data: ADS/JAO.ALMA \#2019.1.00441.S. ALMA is a partnership of ESO (representing its member states),
NSF (USA) and NINS (Japan), together with NRC (Canada), MOST and ASIAA (Taiwan), and KASI (Republic of Korea), in cooperation with the Republic of Chile. The Joint ALMA Observatory is operated by ESO, AUI/NRAO and NAOJ. The authors acknowledge assistance from Allegro, the European ALMA Regional Center node in the Netherlands. \\

M.R. acknowledges the support of the NWO Veni project "Under the lens" (VI.Veni.202.225). M.R. and J.H. acknowledge support of the VIDI research programme with project number 639.042.611, which is (partly) financed by the Netherlands Organisation for Scientific Research (NWO). M.A. acknowledges support from FONDECYT grant 1211951, ``CONICYT + PCI + INSTITUTO MAX PLANCK DE ASTRONOMIA MPG190030'' and ``CONICYT+PCI+REDES 190194''.

\end{acknowledgements}

\bibliography{references.bib}{}
\bibliographystyle{aasjournal}

\end{document}